\documentclass[
reprint,
onecolumn,
nofootinbib,
amsmath,amssymb, aps
]{revtex4-2}
\pdfoutput=1

\DeclareUnicodeCharacter{2009}{\,}
\DeclareUnicodeCharacter{2215}{\,}
\usepackage[utf8]{inputenc}
\usepackage{graphicx}
\usepackage{amsmath}
\usepackage{amssymb}
\usepackage{xcolor}

\begin{document}

\title{Dynamical Dipolar Condensate Finite Temperature Stochastic Gross--Pitaevskii--Boltzmann Model
}
\author{Nick P. Proukakis$^{1}$}
\email[E-mail: ]{nikolaos.proukakis@newcastle.ac.uk}
\author{Gerasimos Rigopoulos$^{1}$}
\email[E-mail: ]{gerasimos.rigopoulos@newcastle.ac.uk}
\author{Alex Soto$^{1}$}
\email[E-mail: ]{alex.soto@newcastle.ac.uk}
\affiliation{$^{1}$ School of Mathematics, Statistics and Physics, Newcastle University, Newcastle upon Tyne, NE1 7RU, UK}

\begin{abstract}
{\noindent 
We formulate a generalized self-consistent stochastic quantum kinetic theory for finite-temperature ultracold Bose gases interacting  via a generic long-range interaction, applicable to a broad range of systems, by means of Keldysh non-equilibrium field theory: such model is explicitly cast in the context of dipolar atomic gases, and is also shown to encompass established stochastic and kinetic treatments for ultracold atomic gases with local interactions as special cases. The condensate and low-lying modes are collectively described by a stochastic Gross-Pitaevskii equation with two collisional terms and their corresponding stochastic noise terms, with thermal particles dynamically modelled through a self-consistently coupled Quantum Boltzmann equation and dipolar interactions included by means of a coupled Poisson-like equation. Additional use of Bogoliubov-de Gennes analysis generating the Lee-Huang-Yang correction term relevant in the $T=0$ quantum-fluctuation-dominated regime, allows us to postulate the extension of such model offering a plausible scheme for interpolating between quantum-dominated and thermal-dominated fluctuation regimes, the consistency of which remains to be tested against experimental observations.

}
\end{abstract}

\maketitle

\section{Introduction}

The aim of this work is to discuss a generalized framework unifying some of the most common approaches for modelling ultracold atomic gases with contact interactions, and extend it to model systems with arbitrary long-range interactions at finite temperatures. While such a model can be generically relevant for a broad range of physical systems, the emphasis in this work is on the modelling of dipolar atomic gases~\cite{Chomaz_2022,Bottcher_2021}, due to the significant experimental impetus in this field (see, e.g.,~\cite{pfau_2005,lev_dysprosium,ferlaino_erbium,pfau_collapse_2008,Kadau_2016,chomaz_droplets,Schmitt_2016,Wenzel_2017,Tanzi_2019,chomaz_prx_2019,bottcher_2019_prx,petter_ferlaino_no_LHY,sohmen_politi_21,Norcia_2021,Tanzi_2021,Casotti_2024,Recati_2023,chomaz2025_review}), and the current lack of a fully self-consistent dynamical theory encompassing both quantum and thermal effects. We believe that this work\footnote{Note that the full model presented in this work was first posted on arXiv in July 2024, accessible as arXiv:2407.20178v1.} makes significant steps in that direction,  offering an extended complementary view to existing approaches for dipolar gases, which primarily focus on the role of quantum fluctuations near the instability regime around $T \approx 0$ where quantum fluctuations (and the so-called Lee-Huang-Yang~\cite{lee_huang_yang} correction to the ground state energy) become important~\cite{fischer_LHY,lima_pelster_11,lima_pelster_12,wachtler_santos,bisset_2013,Bisset_2016,baillie_droplet,ferrier_droplets,Saito_2016,app8101998,chomaz_prx_2019,natale_2019,chomaz_droplets,blakie_2020,Norcia_2021,houwman_2024,kirkby_kz}, 
or $T>0$ settings either in equilibrium~\cite{Ronen_2007,bisset_T_dipolar,blakie_2013,boudjemaa_shlyapnikov,Boudjemaa_2015,baillie_2016,Boudjemaa_2016,Boudjemaa_2017,Aybar_2019,Aybar_2020,boudjemaa_guebli,politi_pohl_23,sanchez-baena-24,maucher_dipolar,he_maucher_2025}, by means of quantum Monte Carlo~\cite{cinti_monte_carlo,boronat_droplet,bombin_2024}, or based on alternative (often simpler) dynamical models~\cite{fischer_LHY,linscott_blakie_14,bisset_2015,blakie_2016,bland_poli_22,poli2024synchronizationrotatingsupersolids,kirkbycomplexlangevin}.
Implications of this model and the quantification of the relevance of the various terms to different experimentally relevant scenarios (including different extent of near- and far-from-equilibrium, and different dimensionalities) remain outside the scope of the present work and are deferred to future work.

This paper is structured as follows:
Sec.~\ref{sec:history} gives a brief historical overview of some of the most common methods used in the field of ultracold atomic gases interacting via the usual s-wave scattering. Sec.~III gives the basic system Hamiltonian, and outlines the key steps in the derivation of our model in the thermal-dominated regime, with our final equations for dipolar condensates$^{1}$ given in Sec.~IV, with corresponding equations in a generic long-range potential given in Appendix~A. Sec.~V shows that in the limit of only s-wave scattering (contact interactions) our equations constitute an extended hybridization of the commonly used Zaremba-Nikuni-Griffin (`ZNG') and Stochastic (Projected) Gross-Pitaevskii Equation (S(P)GPE) established formalisms, clearly highlighting the steps for reducing our theory to such simpler, yet remarkably powerful, models. In Sec.~VI we focus on dipolar atomic gases, comment on the interplay between thermal and quantum fluctuations, and -- by analogy to established models for contact-interacting cold atoms -- we provide reduced sets of equations for dipolar equations valid in appropriate limits. We anticipate that equations of this kind will be extensively used in the following decade to address a number of problems of direct relevance to ultracold dipolar gases. Appendices B and C provide further technical details associated with the operator appearing in our Poisson-like equation for the dipolar interaction and the emergence of the Lee-Huang-Yang contribution to the system's ground state energy.

\section{Brief Historical Overview of Dynamical Bose Gas Theories beyond the GPE \label{sec:history}}

Modelling non-equilibrium quantum gases is a topic with a very rich history. While a significant part of the early literature was focused on obtaining equations for modelling the more strongly-interacting liquid Helium, the emergence of ultracold atomic gases in the 1990's provided an ideal dilute, weakly-interacting, and strongly-controllable system for which {\em ab initio} theories could be directly relevant and compared to. Since then, established approaches~\cite{kadanoff_baym_book,nozieres_pines_book,popov_book,fetter_walecka_book,griffin_book} were revisited and appropriately extended, to take care of the novel features, observations and challenges presented by the dynamical richness introduced by the inhomogeneous external potentials.
Mean-field equations, like the celebrated Gross-Pitaevskii Equation (GPE) first proposed as a suitable phenomenological model for liquid Helium~\cite{gross_1,gross_2,pitaevskii_1,pitaevskii_2}, quickly became the workhorse of weakly-interacting ultracold quantum gases, modelling diverse scenarios~\cite{RevModPhys.71.463,Pitaevskii,Pethick_Smith_2008}, such as linear excitations, sound waves, bright and dark solitons, quantum vortices, Josephson dynamics, persistent currents and superflow, superfluid mixtures, spinor condensates, to mention a few.
The GPE is in fact a remarkably versatile tool, both in terms of modelling near-zero-temperature weakly-interacting quantum gases -- for which the quantum depletion is small -- and also (perhaps somewhat counter-intuitively) the highly-populated modes of a finite-temperature system as an effective field theory~\cite{svistunov_91,kagan_svistunov_94,kagan_svistunov_97,berloff_svistunov_02,davis_morgan_02,Brewczyk_2007}. Such a model -- which is the simplest classical description based on a (non-relativistic) binary interaction Hamiltonian -- gradually evolved into more extended models, such as the `Zaremba-Nikuni-Griffin' (ZNG) and the Stochastic (Projected) Gross-Pitaevskii Equation (SPGPE), as alternative models of the full system Hamiltonian -- see, for example the recent reviews
~\cite{Proukakis13Quantum,proukakis_jackson_08,blakie_bradley_08,berloff_brachet_14,stoof_99,griffin_nikuni_zaremba_2009,gardiner2017quantum,chantesana_kinetic_2019}.

Beyond the addition of a phenomenological dissipation term first proposed by Pitaevskii~\cite{pitaevskii_3,pitaevskii_4} which is being commonly applied as a first approximation for finite-temperature effects (see, e.g.,~\cite{choi_morgan_98,tsubota_kasamatsu_03}), such finite-temperature approaches can be loosely speaking classified into a few rather different categories briefly summarized below.
We feel it is useful to do so here, in order to motivate the different limits of dipolar gas theories proposed in Sec.~VI, but note that we do not attempt to give a fully comprehensive overview of all active approaches to date, for which we refer the reader to an earlier edited volume~\cite{Proukakis13Quantum}, related review articles~\cite{proukakis_jackson_08,blakie_bradley_08,berloff_brachet_14} and works cited therein, or citing such works.

The first broadly applicable method we wish to highlight here is based on techniques borrowed from the modelling of macroscopic quantum liquids, based on what might be called a `two-gas' approximation, in direct analogy to the very successful `two-fluid' model of liquid Helium~\cite{tisza_38,landau_41}. In the `two-gas' model, one starts by directly invoking symmetry breaking of the Bose field operator, treating the beyond-condensate contribution perturbatively, under appropriate approximations. Such a methodology, first developed by Kirkpatrick and Dorfman~\cite{kirkpatrick_dorfman_83,kirkpatrick_dorfman_85a,kirkpatrick_dorfman_85b,kirkpatrick_dorfman_85c} (see also Eckern~\cite{eckern_ulrich_84}) was significantly advanced by Zaremba, Nikuni and Griffin (`ZNG')~\cite{zaremba_nikuni_99,griffin_nikuni_zaremba_2009} and leads to a dissipative finite-temperature GPE self-consistently coupled to a quantum Boltzmann equation for the thermal cloud: such an approach has been successfully applied~\cite{ZNG3} to study damping of collective modes~\cite{ZNG4,ZNG-scissors,Straatsma_2016}, solitons~\cite{ZNG-soliton}, vortices~\cite{vortexT1,vortexT2,vortexT3}, Josephson junction dynamics~\cite{Xhani20,xhani_proukakis_22} and mixtures~\cite{zng_mixture_1,zng_mixture_2,ev2,lee_jorgensen_16,lee_jorgensen_18}. While an excellent model for a broad temperature range, and ideal for accurate treatment of collective modes, this model cannot describe the presence of large fluctuations, such as those arising in low-dimensional systems and near the critical region. As such, this model cannot describe the physics of phase transitions, although -- rather remarkably -- it can in fact capture fairly well quasi-adiabatic condensate growth once a well-formed small condensate (mimicked by an artifical initial seed) has formed~\cite{ZNG-growth,ev1}. 
Another important shortcoming of the ZNG method manifesting itself in some regimes, and ultimately limiting its validity, is that it only accounts for single-particle `Hartree-Fock' energies dressed by the potential and interactions, in the so-called `Popov' approximation ignoring anomalous pair correlations~\cite{griffin_96}, but it does not include {\em quasiparticle} physics associated with the Bogoliubov transformation~\cite{bogoliubov_47} mixing single-particle creation and annihilation operators. This point will become relevant later on, when discussing the equations for modelling dipolar gases in the quantum droplet regime, when such corrections become critical~\cite{lee_huang_57}.

Various other closely-related approaches were developed in parallel to ZNG~\cite{proukakis_burnett_96,proukakis_thesis_97,proukakis_burnett_98,proukakis_morgan_98,proukakis_01,morgan_00,davis_jpb_01,walser_williams_99,walser_cooper_01,wachter_walser_01a,wachter_walser_01b}, aiming to extend beyond such an approximation through the inclusion of so-called anomalous averages (see, e.g., their review in Ref.~\cite{proukakis_jackson_08}), but -- despite limited applications~\cite{walser_04} -- these did not directly lead to broadly implemented numerical modelling (but see also discussion below).
Nonetheless, we specifically highlight here the so-called (full) Hartree-Fock-Bogoliubov (HFB) equations, in which the normal and anomalous averages are defined by means of the coherence factors appearing within the Bogoliubov-de Gennes equations~\cite{proukakis_jackson_08}: despite some problems associated with the decades-old discussion about conserving versus gapless approximations~\cite{hohenberg_martin,kadanoff_baym_book,proukakis_01,wachter_walser_01a,wachter_walser_01b}, the latter approach may have some benefits in our present context, as inclusion of Bogoliubov mixing actually encapsulates and can thus generate the condensate quantum depletion, the latter known to be critical to the dipolar droplet and supersolid regimes.

A model consistent with ZNG can, in fact, be extended beyond the symmetry breaking assumption, by explicitly preserving the operator nature of the condensate mode, in what are known as number-conserving approaches: first introduced in~\cite{girardeau_arnowitt_59,girardeau_98}, such approaches have in fact led to a somewhat cumbersome set of equations~\cite{gardiner_97,morgan_00,gardiner_morgan_07,billam_2013} which has successfully modelled early collective mode experiments with thermal dissipation~\cite{morgan_2003,morgan_05}. We also note that an alternative set of equations preserving phase fluctuations to all orders, and thus fully describing quantum depletion and quantum effects, can also be obtained~\cite{Petrov00_1D,Petrov00Bose,Petrov01Phase,Stoof02Low,Stoof02Erratum}, but such a model has only been applied to date to a small number of equilibrium settings~\cite{Stoof02Low,Stoof03Dimensional,proukakis_06b,cockburn_negretti_11,Henkel_2017}.

A very different class of approaches is based on the widely-accepted notion of effective field theories. The idea is to split the entire set of modes of the Bose gas to a set of modes encompassing, on the one hand, cumulatively the condensate and all the modes whose dispersion relation is significantly affected by the presence of a condensate, and, on the other hand, the higher-lying modes which are effectively thermal. This splits the system modes into a low-lying `classical field' or `coherent' region, and the higher-lying modes (above a certain energy cutoff) into an `incoherent' region. In the most general setting, both of these system sub-parts obey appropriate dynamical equations. 
Such an approach was first developed by Stoof in a series of papers based on a formulation in terms of coherent states utilizing the non-equilibrium Keldysh formalism~\cite{stoof_97,stoof_chapter_98,stoof_99}. Although the most general formulation by Stoof also includes an explicitly dynamical thermal cloud obeying a quantum Boltzmann equation, such coupling was not numerically implemented, with the thermal cloud treated instead as a heat bath with fixed temperature and chemical potential.
This is a valid approximation in the usual regime where the dominant dynamics occurs within the highly-populated low-lying modes of the system, 
which can therefore be effectively described as a classical field modelled by a dissipative GPE with stochastic noise term(s). Such a nonlinear Langevin equation is known as the Stochastic  Gross-Pitaevskii Equation (SGPE)\footnote{This is sometimes also referred to as a time-dependent stochastic Ginzburg-Landau equation~\cite{berloff_brachet_14}.}. These considerations (and approximations) led to the first numerical application of such a SGPE in ultracold gases~\cite{stoof2001dynamics}, in the context of qualitatively modelling a reversible condensate formation process (induced by cycling through the phase transition~\cite{ketterle_98}).

Parallel to the above developments, Gardiner and Zoller developed a very effective quantum kinetic theory~\cite{jaksch_gardiner_97,gardiner_zoller_98,davis_gardiner_2000} based on quantum optics approaches, which led to the first modelling of condensate growth experiments~\cite{davis_growth_1997,davis_growth_1998,davis_growth_2002}, and was later also cast into an extended stochastic model~\cite{Gardiner_2002a,Gardiner_2002b,Bradley08Bose}. Although derived by means of very different formalisms, the qualitative similarity of the approximations made led to a stochastic approach closely related to the SGPE of Stoof: the main physical difference (besides the absence of explicit incoherent region dynamical equations) is that the full stochastic projected GPE (SPGPE) equation also includes an extra collisional `scattering' process with its own associated additional noise term, beyond the dissipative and noise contributions already included in Stoof's treatment:
Such an additional contribution has enabled Bradley {\em et al.} to discuss the distinction between {\em number} and {\em energy} damping mechanisms (see, e.g.,~\cite{rooney_collective_12,rooney_allen_16,bradley_spgpe_scipost_20,mehdi_hope_23}).
Moreover, it is important to highlight the explicit inclusion of a projector in the SPGPE, required numerically to ensure the two `subspaces' of coherent and incoherent regions remain well-separated during numerical calculations and thus avoid aliaising\footnote{See also Refs.~\cite{davis_gardiner_2000,blakie_bradley_08,gardiner2017quantum}) for a more detailed discussion of those, and other, differences between such approaches.}. 

Although the coupling to a quantum Boltzmann equation has already been discussed in~\cite{stoof_99,Duine_2001}(see also~\cite{proukakis_rigopoulos_prd_23,proukakis_rigopoulos_arxiv_24}), all numerical implementations of the various stochastic equations to date have made the explicit assumption of a large thermal cloud which remains essentially unperturbed during the evolution, and can thus act as a heat bath providing the dissipation and stochastic noise terms for the effective low-lying coherent region. In so doing, one is implicitly forced to drop the physically-exact relaxation requirement to a Bose-Einstein distribution, which instead becomes replaced by a Rayleigh-Jeans distribution [but see also Ref.~\cite{deuar_2017}], an excellent approximation for all highly-occupied modes (below the chosen cutoff), with typically many particles per mode, and at the characteristic temperatures studied~\cite{blakie_bradley_08}.
Beyond Stoof's pioneering reversible formation work~\cite{stoof2001dynamics},
early dynamical studies on continuous atom laser operation~\cite{Proukakis03Coherence} (revisited in ~\cite{lee_haine_15}), 
and rotating condensates~\cite{Bradley08Bose}, and a plethora of equilibrium studies~\cite{Stoof02Low,proukakis_06b,cockburn_gallucci_11,cockburn_negretti_11,gallucci_cockburn_2012,Cockburn12Ab,Garrett_2013,Henkel_2017}, this approach has been used to provide new insight into non-equilibrium phenomena in ultracold atomic gases, including most notably
spontaneous defect generation during dynamical growth in the context of  non-equilibrium phase transition physics and Kibble-Zurek universality~\cite{Weiler08Spontaneous,Proukakis09The,Damski10Soliton,Su13Kibble,Rooney13Persistent,Kobayashi16Thermal,Kobayashi16Quench,Liu18Dynamical,Liu20Kibble,bland_marolleau_20}, and the subsequent relaxation process~\cite{Comaron19Quench,groszek_comaron_PRR,brown_bland_21}, soliton~\cite{cockburn_nistazakis_10,cockburn_nistazakis_11,spgpe_soliton_2011}, vortex~\cite{rooney_bradley_10,rooney_allen_16,mehdi_hope_23} and persistent current~\cite{Rooney13Persistent,bland_marolleau_20,mehdi_bradley_21} dynamics. Other works include studies of collective mode dynamics~\cite{rooney_collective_12,Straatsma_2016,bradley_spgpe_scipost_20}, condensate mixtures~\cite{bradley_blakie_14,liu_stoch_mixtures_16,Roy_2021,Roy_2023}, and sound propagation~\cite{Ota18Collisionless}.
Some subtleties associated with the above-mentioned approximations, including on the role of the ensemble~\cite{deuar_2017}, atom number conservation, and a different hybrid handling of the noise term which maintains a fully quantum description of the low-energy fields using the positive-P representation (which could prove crucial in the case of finite-temperature dipolar droplets) can be found in Refs.~\cite{deuar_gangardt_09,swislocki_deuar_2016,deuar_2017,deuar_2019}. See also Ref.~\cite{bradley_low_D_spgpe,keepfer_liu_22} for other implementation in lower dimensions (and the somewhat related work~\cite{thomas_davis_relaxation}.

At this point it is also important to mention that, as the lowest-order equation for a classical field~\cite{polkovnikov_03,polkovnikov_10}, the Gross-Pitaevskii equation can in fact also describe  a pathway to equilibration at fixed energy and particle number -- both for the condensate and for the highly-populated, low-lying, non-condensate modes -- as already discussed in~\cite{svistunov_91,kagan_svistunov_94,berloff_svistunov_02} -- with this point made more rigorous upon the explicit addition of a projector, in what is known as the Projected Gross-Pitaevskii Equation (or PGPE)~\cite{davis_morgan_01,davis_jpb_01,davis_morgan_02} (see also related discussion in Ref.~\cite{Brewczyk_2007}). Relaxation to a semi-classical equilibrium state can be achieved either by appropriately seeding the initial conditions with stochastically-sampled quantum fluctuation contributions~\cite{steel_olsen_98,sinatra_jmo_00,sinatra_prl_01,Sinatra_2002} (usually referred to as truncated Wigner approximation), also extended to include temperature effects at high condensate fraction situations, or by simply seeding the initial condition with populated modes up to a cutoff, with such modes having random phases~\cite{berloff_svistunov_02}. Due to the mixing facilitated by the nonlinear interaction term in the GPE, any such initial seeding will result in the correct quasi-classical equilibrium distribution for a given particle number, temperature and total system energy. 
A more controllable initial condition to be used as input for dynamical GPE propagation in the context of a finite-temperature system can also be obtained by means of dynamical equilibration of the S(P)GPE, with particle number and temperature respectively fixed by the imposed bath chemical potential and temperature, as first implemented in~\cite{proukakis_schmiedmayer_06}. Irrespective of the details of setting up the initial conditions, such a `classical field' method will equilibrate to the correct distribution, but under the assumption of validity of Rayleigh-Jeans, and this method has also been used to study some dynamical relaxation.
Early applications of these various classical field method implementations can be found in Refs.~\cite{blakie_bradley_08,Proukakis13Quantum}.

As stated earlier, the primary aim of this introduction was simply to cast the discussion within this paper in the broader context of key preceeding quantum kinetic theory advances over the last few decades, such that our presented findings can be easily compared against such models.  As such, we are aware of not having done full justice to other exciting methodological developments in the community, and we sincerely apologise for any unintentional omissions of alternative theories\footnote{For completeness, and while not directly related to our present discussion despite their close connections, we also note here the existence of other approaches: most notably, the `raw' truncated Wigner for the whole field~\cite{steel_olsen_98,Sinatra_2002,norrie_2005}, the positive P-representation of quantum optics~\cite{drummond_gardiner_positive_P} for which the quantum operators $\hat{\Psi}$ and $\hat{\Psi}^{\dag}$ are treated via stochastic equations for two independent classical fields $\phi$ and $\phi^*$~\cite{steel_olsen_98,deuar_drummond_positive_P}, based on an underlying Fokker-Planck equation describing the positive-P quasiprobability phase space distribution function obtained from a master-equation approach; the field-theoretic two-particle-irreducible (2PI) formalism~\cite{calzetta_hu_book,rey_2PI,gasenzer_2PI_a,gasenzer_2PI_b}; the stochastic wavefunction approach~\cite{breuer_book}.}, or their implementations.

Although the GPE has now become the workhorse model for weakly-interacting quantum gases, one should not forget that it was initially proposed for modelling liquid helium which is not well-described by simple contact interactions; in fact, a more accurate model of the helium interactions is based on a non-local interaction potential~\cite{berloff_99,berloff_roberts_99,berloff_brachet_14}. 
Although the familiar realm of dilute weakly-interacting ultracold atomic gases~\cite{RevModPhys.71.463} is based on interactions which -- under these conditions -- can be effectively modelled as s-wave scattering, and thus a contact potential with an effective interaction strength set by the s-wave scattering length, the gallery of available quantum gases is constantly being broadened.
In this context, we note the increasing interest in systems with non-negligible magnetic dipole moments, recently reviewed in Ref.~\cite{Chomaz_2022}: these range from the well-studied dipolar condensates, which feature a combination of short-range (s-wave) and long-range (dipolar) interactions, to more elaborate experimentally-accessible systems, such as Rydberg atoms, or Bose-Einstein condensates of dipolar molecules, where particularly exciting progress has been made very recently~\cite{Bigagli_2024}. With that in mind, the present work introduces in a very general manner a set of self-consistent finite-temperature equations describing partially-condensed systems exhibiting both short- and (arbitrary) long-range interactions, in a manner which  includes both dissipative and stochastic noise terms, alongside self-consistently coupled Boltzmann equations for the incoherent particles and Poisson-like equations modelling the (arbitrary) long-range interaction.
In the thermal-fluctuation-dominated regimes, such equations are more general than all previous treatments to date. Nonetheless -- as we discuss below -- the nature of our approximations is such that quantum fluctuations are not self-consistently treated to {\em all} orders, and the implications of -- and potential remedies for -- such an issue -- are also touched upon within this manuscript.


In brief, this work is based on a standard Keldysh path integral formalism~\cite{kamenev_book}, which -- using the tools discussed in detail in our companion paper~\cite{proukakis_rigopoulos_arxiv_24} --
leads to a self-consistent finite-temperature stochastic quantum model for ultracold Bose gases in the most general scenario of a co-existing short and long-range interaction between the particles, such as that found in dipolar gases.
Our model thus generalizes beyond the above-discussed quantum gas models, as highlighted (where appropriate) within the manuscript.

Sec~III briefly summarizes -- following closely our related derivation in the context of gravitational coupling~\cite{proukakis_rigopoulos_prd_23,proukakis_rigopoulos_arxiv_24} -- the key steps in the derivation of our main equations, which are presented for clarity in Sec.~IV. Secs.~V and VI then explicitly relate such generalized equations to a broad range of applicable cold-atom theories discussed in the literature, in the specific contexts of local and dipolar interactions respectively. Our findings are briefly summarized in a final discussion section.
Appendix A gives our full equations in a more general potential form, such that their mathematical expression can be readily available in the future for any more general non-local interaction potential becoming relevant. Appendix B gives some further context to the non-local operators appearing in the main manuscript, while Appendix C provides details on the inclusion of quantum effects to all orders in the near zero-temperature limit.

\section{System Hamiltonian and Effective Action for a Dipolar Atomic Gas}


\subsection{Basic Hamiltonian -- Effective Action}

Our starting point is the usual binary interaction hamiltonian given (in natural units $\hbar = k_B = 1$) by
\begin{equation}
\label{hamiltoniangen}
H = \int d^3 \mathbf{r} \left( - \frac{1}{2 m} \psi^{\ast} \nabla^2 \psi +
V_{ext}\psi^\ast \psi \right) + \frac{1}{2} \int d^3 \mathbf{r} \int d^3 \mathbf{r}'  \psi^\ast(t,\mathbf{r}) \psi^\ast(t,\mathbf{r}') U (\mathbf{r}, \mathbf{r}') \psi(t,\mathbf{r}')\psi(t,\mathbf{r}) \;.
\end{equation}
Here $V_{ext}$ is the external potential and the  term $U(\mathbf{r},\mathbf{r}')$ describes the bosonic interaction potential between two particles.
In this work, we assume that the general potential $U (\mathbf{r} - \mathbf{r}') $ can be cast in terms of two contributions, a local one ($U_{l}$) expressible in terms of the usual contact interaction term (of effective strength $g$)  and a non-local one ($U_{nl}$), with $U (\mathbf{r} - \mathbf{r}') $ thus decomposable in the form
\begin{equation}
U (\mathbf{r} - \mathbf{r}') =
U_{l} (\mathbf{r} - \mathbf{r}') + U_{nl} (\mathbf{r} - \mathbf{r}') =
g \delta (\mathbf{r} - \mathbf{r}') +
U_{nl} (\mathbf{r} - \mathbf{r}') \;. \label{U-s-nl}
\end{equation}
The first term corresponds to the typical contact interaction with coupling $g=4 \pi a/m$, where $a$ denotes the relevant s-wave scattering length, and is valid for sufficiently dilute low-temperature gases (i.e.~the limit of weak interactions, and small diluteness parameter $n a^3 \ll 1$, where $n$ labels the atomic density). 

A general non-local interaction contribution, $U_{nl} (\mathbf{r} - \mathbf{r}')$, can arise in many distinct scenarios, including -- for example -- dipolar interactions, or Rydberg interactions, with
details of the particular nature of the arising interaction in each case found in Ref.~\cite{Chomaz_2022}.
Other systems where similar non-local interactions are relevant, include
indirect exciton studies~(see, e.g.,~\cite{conti_biexcitons}), or dipolar molecules~\cite{Bigagli_2024,dipolar_molecule_theory}.
The theory presented in this work is directly applicable for any type of long-range interaction of the form of Eq.~(\ref{U-s-nl}), provided we can formally write down a matrix inversion for $U_{nl}$ (see Appendix A).

Although the more general final form of equations for a general potential are given explicitly in Appendix A, in the remainder of this paper we have chosen to state our results  explicitly in terms of a long-range interaction written in the form of the usual long-range dipolar gas interatomic potential form~\cite{li_you_2000,goral_dipolar,santos_2000,li_you_2001,blume_2006}.
This is in order to connect here directly to the community of
 dipolar atomic condensates, for which the form of the interatomic potential is very well known.
 With that in mind, we set 
$U_{nl} (\mathbf{r} - \mathbf{r}') \rightarrow U_{dd} (\mathbf{r} - \mathbf{r}')$, or
%
\begin{equation}
U_{nl} (\mathbf{r}) \rightarrow U_{dd} (\mathbf{r}) = \frac{C_{dd}}{4 \pi r^3} \left( 1 - 3
\frac{(\mathbf{n} \cdot \mathbf{r})^2}{r^2} \right) \;, \label{U-dd}
\end{equation}
where the unitary vector $\mathbf{n}$ is the dipolar axis, which describes the direction of orientation of the dipoles and this interaction is controlled by the parameter $C_{dd}$, whose value is set by~\cite{Chomaz_2022} (i) $C_{dd} = \mu_0m^2$ for magnetic dipoles (where $\mu_0$ is the magnetic permeability in vacuum and $m$ the magnetic dipole moment), or (ii) $C_{dd} = d^2/\epsilon_0$ for electric dipoles (where $\epsilon_0$ is the vacuum permittivity).

Using the relation
\begin{equation}
\frac{1}{r^3} \left( 1 - \frac{3 (\mathbf{n} \cdot \mathbf{r})^2}{r^2}
\right) = - \frac{4 \pi}{3} \delta (\mathbf{r}) - (\mathbf{n} \cdot
\nabla)^2 \left( \frac{1}{r} \right) \;,
\end{equation}
 the full non-local potential of Eq.~\eqref{U-s-nl}  can be recast into the form~\cite{giovanazzi_2002,odell_2004,odell_2005,Bao_2010}
\begin{equation}
U (\mathbf{r} - \mathbf{r}') = \left( g - \frac{C_{dd}}{3} \right) \delta (\mathbf{r} - \mathbf{r}') 
- C_{dd}  \frac{1}{4 \pi} (\mathbf{n} \cdot \nabla)^2 \left(
\frac{1}{r} \right) \;,
\end{equation}
where one part of the dipolar interaction plays the role of renormalizing the strength of the local interaction.

Inserting this form into the Hamiltonian we can write it as 
\begin{eqnarray}
\label{gpe2}
H &=& \frac{1}{2} \int d^3 \mathbf{r} \left( - \frac{1}{2 m}
\psi^{\ast} (\mathbf{r}, t) \nabla^2 \psi (\mathbf{r}, t) + V_{ext}
| \psi (\mathbf{r}, t) |^2 + \frac{1}{2} \left( g - \frac{C_{dd}}{3}
\right) | \psi (\mathbf{r}, t) |^4 \right) \nonumber\\
& & - \frac{1}{2}
\frac{C_{dd}}{4 \pi} \int d^3 \mathbf{r} \int d^3 \mathbf{r}'
\psi^{\ast} (\mathbf{r}, t) \psi^{\ast} (\mathbf{r}', t) \left(
(\mathbf{n} \cdot \nabla)^2 \frac{1}{| \mathbf{r} - \mathbf{r}' |}
\right) \psi (\mathbf{r}', t) \psi (\mathbf{r}, t) \;.
\end{eqnarray}


In order to proceed with our derivation, we now consider\footnote{This procedure is motivated by our closely related cosmological work~\cite{proukakis_rigopoulos_prd_23,proukakis_rigopoulos_arxiv_24}, which derived a general set of self-consistent equations focussing on a gravitational (instead of dipolar) long-range interaction.} the corresponding action for a bosonic particle $\psi (t,\mathbf{r})$ with mass $m$, defined as ($c=1$)
\begin{equation}
S = \int d t \bigg(i \int d^3 \mathbf{r} \ \psi^{\ast} \dot{\psi}- H \bigg) \;.
\end{equation}
Via a Hubbard-Stratonovich transformation we introduce an auxiliary field $V$ and recast the above Hamiltonian, mapping our problem to that of an equivalent effective action\footnote{Note that we have chosen to apply the Hubbard-Stratonovich transformation to the so-called direct channel; the possible ramifications of a different choice~\cite{Altland_Simons_2010}, in conjunction with an appropriate fast-slow field split (discussed below) are left to future work. We thank Axel Pelster for raising the point with us.}
\begin{eqnarray}
\label{actiondip}
S &=& \int d t d^3 \mathbf{r} \bigg[ i \psi^{\ast} \dot{\psi} + \frac{1}{2
m} \psi^{\ast} \nabla^2 \psi - V_{ext} \psi^{\ast}\psi - \frac{1}{2} \left(
g - \frac{C_{dd}}{3} \right) (\psi^{\ast}\psi)^2 \nonumber\\
& & \hspace{1.5cm} - C_{dd} V \psi^{\ast}\psi +
\frac{C_{dd}}{2} V \frac{\nabla^2}{(\mathbf{n} \cdot \nabla)^2} V \bigg] \;.
\end{eqnarray} 
Apart from the time derivative term, the object in brackets is effectively equivalent to the hamiltonian in \eqref{gpe2}. To see this, it is enough to take the equation of motion for $V$ (obtained by varying the action \eqref{actiondip} with respect to $V$) and plug it back into the action, resulting in the original hamiltonian. More details of such equivalence are given in Appendix~A, cast (for generality) in the more general context of a general operator.

Having recast the problem in terms of the above effective action \eqref{actiondip}, we can now directly capitalize on our companion works~\cite{proukakis_rigopoulos_prd_23,proukakis_rigopoulos_arxiv_24} deriving corresponding equations in the context of locally-interacting bosonic particles additionally bound by the gravitational potential, which plays the role of a long-range interaction.
To avoid duplication, and in the interest of clarity, we focus here directly on the implications of our model for ultracold atomic gases, without reproducing all the lengthy technical details of the derivation in this manuscript, referring  instead the reader to those works~\cite{proukakis_rigopoulos_prd_23,proukakis_rigopoulos_arxiv_24}.
Instead we only give here a brief overview of the key steps, highlighting the approximations made, which enable us to understand the limitations of our self-consistently obtained final set of equations.

\subsection{A Summary of Key Steps:}
A summary of the steps undertaken to derive our equations is given below -- for more details readers may consult our companion work~\cite{proukakis_rigopoulos_arxiv_24}:
\begin{enumerate}

\item {\bf Set up the ``Classical'' and ``Quantum'' fields of the Schwinger-Keldysh formalism:}\\
We employ the Schwinger-Keldysh formalism in our action \eqref{actiondip}, such that the evolution of a non-equilibrium density matrix involves a doubling of the Hilbert space: specifically, we write the Schwinger-Keldysh action as $S_{\rm SK} = S[\psi^{+},V^{+}]-S[\psi^{-},V^{-}]$, where the superscript $+$ defines the fields evolving forward time and the superscript $-$ the fields evolving on the backward time contour. As the integration is done backwards in the second term, such term is subtracted.
Then, we use the Keldysh rotation, to construct so-called `classical' ($\psi$, $V_d$, left entries) and `quantum' ($\psi_q$, $V_d^q$, right entries) terms via the sums and differences across the two contours, in the form
\begin{eqnarray}
& &\psi = \frac{1}{\sqrt{2}} (\psi^+ + \psi^-), \quad \psi^q = \frac{1}{\sqrt{2}} (\psi^+ - \psi^-) \nonumber\\
& & V_d = \frac{1}{2} (V^+ + V^-), \quad V_d^q = \frac{1}{2} (V^+ - V^-) \nonumber
\end{eqnarray}
Note that the stochastic terms ultimately appearing in our equations stem from consideration of these quantum fields $(\psi^q, V_d^q)$ to second order.

\item {\bf Split the bosonic field in slow and fast components:}\\
We split both `classical' and `quantum' components into two parts, through:
\begin{equation}
\psi=\Phi_0+\varphi
\hspace{1.0cm} {\rm and} \hspace{1.0cm}
\psi^q=\Phi^q+\varphi^q \;.
\end{equation}
with the first terms corresponding to the slow (or low-momentum) component, and the second part to the fast (or higher-momentum) part, noting that the location of such a split is arbitrary. Thus, the slow component $\Phi$ 
would typically contain elements with a higher order of coherence, but not (necessarily) fully condensed, relegating the incoherent part to $\varphi$. 
In the case of the auxiliary potential, we will work considering that $V_d$ and $V_d^q$ are slow quantities\footnote{This appears to be a sensible prescription, although we note that it is not {\em a priori} guaranteed to correctly encompass all relevant diagrams, something which will be checked in future work.}.
Here we are ignoring possible fast contributions for the long-range potential, meaning that we are only taking in account (macroscopic) mean-field effects of this potential. Doing this, terms in the action with only one $\varphi$ term are discarded due to violation of energy-momentum conservation. This class
of objects are: terms with one slow field $\Phi$ and one fast $\varphi$, terms with three slow $\Phi$ and one fast $\varphi$, and terms with
one slow $\Phi$, one slow $V$ and one fast $\varphi$.
Also, within this work, we will keep up to order two the components $\Phi^q$ and $V_d^q$, while simultaneously maintaining all orders of $\varphi$ (since we will integrate out this field).

\item {\bf Integrate out the fast $\varphi$-particles in the generating functional to get an effective action for the slow fields:}\\
We use the generating functional $Z = \int \mathcal{D} [\Phi V \varphi] e^{i S_{SK}}$, working in perturbation theory, expanding up to second order in the couplings related with the self interaction (i.e. $g-\frac{C_{dd}}{3}$) and the interaction with the auxiliary field $V$ (i.e. $C_{dd}$). Then, we integrate out the fast component $\varphi$ such that we are left with an effective action for the slow parts. In the process of integration we obtain lengthy expressions containing propagators and four-point functions of $\varphi$. In order to manage them, we make use of Wick's Theorem and we discard\footnote{Discarding such terms at this stage is equivalent to setting the coherence factors $u_k=1$ and $v_k=0$ in the Bogoliubov transformation, as discussed in~\cite{proukakis_rigopoulos_prd_23}. See, however, Sec.VI for an alternative way of including such contributions and their potential relevance for dipolar gases.} `anomalous' terms like $\langle\varphi\varphi\rangle$, $\langle\varphi^\ast\varphi^\ast\rangle$, $\langle\varphi^q\varphi^q\rangle$ and $\langle\varphi^{q\ast}\varphi^{q\ast}\rangle$ in a manner closely related to the `Popov' approximation~\cite{griffin_96,griffin_nikuni_book_09,yukalov_kleinert_06}.
After integration, via a Hubbard-Stratonovich transformation we change the terms containing  $(V_d^q)^2$ and $(\Phi^q)^2$ to objects linear in these fields by introducing auxiliary fields $\xi_1$ and $\xi_2$, which will be the noise terms appearing in our equations. Thus, we obtain an effective action for the slow fields, from which we compute the Euler-Lagrange equations after varying with respect to $\Phi^q$ and $V_d^q$, arriving at the equations of motion for the coherent part $\Phi_0$ and the field $V_d$ respectively.

\item {\bf Use of propagator relations and Wigner transforms for the fast component equation:}\\
To recover the dynamics of the fast part, integrated out in the generating functional, we use the Schwinger-Dyson equation for the propagator in the fast fields. We focus on the Keldysh component of the propagator, and taking a Wigner transform, we arrive to the Boltzmann-like equation for the incoherent particles. In the process of taking the Wigner transform, we consider that the mean-field potential $V_{nc}$ (see below) is a slow varying quantity. Thus, the field $V_d$ is slow, as we assumed in the beginning and so are also the number densities of the coherent and incoherent parts, defined to be $n_c=|\Phi_0|^2$ and $\tilde{n} = {\sum}_k f$, where $f$ in the distribution function appearing in the Boltzmann equation. Finally, we pass any sum to an integral considering a continuum limit and we obtain our main equations.


%
\item{\bf Formulation in terms of dipolar interactions:}\\
A direct comparison of the effective action given by Eq.~\eqref{actiondip} 
with the effective action discussed in our companion manuscript in the presence of a gravitational field (Eq.~(4) in Proukakis {\em et al.}~\cite{proukakis_rigopoulos_arxiv_24}), shows their direct correspondence upon carrying out the following redefinitions
\begin{eqnarray}
 g &\rightarrow& \left( g - \frac{C_{dd}}{3} \right) \;, \\ 
   \frac{1}{8 \pi G} \nabla^2 &\rightarrow& \frac{C_{dd}}{2} \frac{\nabla^2}{(\mathbf{n} \cdot \nabla)^2} \;.
\end{eqnarray}
With this in mind, we can now directly write down the final equations relevant for the context of ultracold dipolar atomic gases.

Although our emphasis here is on the experimentally-relevant cold atomic gases, we note that our mathematical approach is in fact more general, and can handle -- at least in principle -- any type of interaction potential, so our methodology might be useful to a broader class of systems. Details of such more general formulation, without resorting to any particular form of the interaction potential, are presented in Appendix~A for any non-local interaction which can be cast in the form of an inverse operator $\cal{O}$$^{-1}({\bf r},\,{\bf r'})$, as such formulation enables us to directly import our developed methodology to such a setting.

\end{enumerate}

\section{Generalized Self-Consistent Stochastic Quantum Modelling for Dipolar Gases}

We can now directly write down the final set of equations arising through the above procedure. They include an equation for the low-lying modes of the system, $\Phi_0$, self-consistently coupled to a quantum Boltzmann equation for the high-lying thermal modes, and a Poisson-like equation for the (main) non-local part of the interaction, constituting (as we show below) a direct hybridization and extension of classical field and kinetic methods extensively used for modelling ultracold atomic gases interacting via contact interactions.

More specifically:
%
\begin{itemize}
    \item (i)
The classical, or `coherent', field $\Phi_0$ obeys a {\bf stochastic Gross-Pitaevskii equation} in the form
\begin{eqnarray}
& & i \frac{\partial \Phi_0 (x)}{\partial t} = \left( - \frac{1}{2 m} \nabla^2 +
V_{ext} + V_c \right) \Phi_0 (x) \label{sgpe-full-1} \\
& & \hspace{1.5cm} - i R \Phi_0 (x) + \xi_1 \label{sgpe-full-2} \\
& & \hspace{1.5cm} - 2 \left( g - \frac{C_{dd}}{3} \right) \int d^4 x' \Pi^R (x', x)
V_{nc} (x') \Phi_0 (x) + \left( g - \frac{C_{dd}}{3} \right)\Phi_0 \, \xi_2 \label{sgpe-full-3} \,, \end{eqnarray}
describing the low-lying highly-populated modes of the system.
This equation features  two collisional terms (identified by the `dissipative' $R$ and `scattering' $\Pi^R$) and their two corresponding complex additive ($\xi_1$) and real multiplicative ($\xi_2$) stochastic contributions [defined respectively in subsequent Eqs.~\eqref{rdef2}, \eqref{Pi-R}, \eqref{realcorrelators-a}, \eqref{realcorrelators-b}]. 
Note that  theoretical studies of dipolar condensates  typically also include (particularly near the instability threshold) an additional term to the usual Gross-Pitaevskii equation [Eq.~(\ref{sgpe-full-1})] in the form
\begin{equation}
i \frac{\partial \Phi_0}{\partial t} = \cdots + \gamma_{QF} |\Phi_0|^3 \Phi_0 \;, \label{extra} 
\end{equation}
where $\gamma_{QF}$ [defined subsequently in Eq.~(\ref{gamma-qf})] is a function of the diluteness parameter $n a^3$~\cite{lima_pelster_11,lima_pelster_12,fischer_LHY,wachtler_santos,baillie_droplet,Bisset_2016} which, more generally, is also temperature-dependent~\cite{Boudjemaa_2016,Boudjemaa_2017,Aybar_2019,maucher_dipolar}.
Such a contribution, known as the Lee-Huang-Yang (LHY) contribution~\cite{lee_huang_57,lima_pelster_11,lima_pelster_12,fischer_LHY,wachtler_santos,maucher_dipolar,petrov_LHY} provides a correction
relevant for reproducing the ground state system energy.
Although it is generally reasonable to ignore such a term for gases interacting via weak effective local (contact) interactions in the rather dilute regime $n a^3 \ll 1$, long-range interactions can, under certain experimentally-accessible regimes, enhance the importance of this term.
Such a term is known to play a key role when mean-field interactions are suppressed such that quantum fluctuations become the dominant contribution, which has been well-studied  in the context of
stabilization of self-bound droplets~\cite{ferrier_droplets,chomaz_droplets},
 with the inclusion of such a term into the usual GPE equation having become the norm in the field of dipolar condensates in recent years (A more extended list of references can be found in the recent review~\cite{Chomaz_2022}).
This contribution  arises naturally
upon linearizing the condensate wavefunction by means of the usual Bogoliubov-de Gennes analysis, but does not directly emerge from our above analysis due to our choice of setting $u_k=1$ and $v_k=0$ discussed in the previous section. Despite the significant success of such `quantum' term in modelling droplets and supersolids in the $T=0$ limit~\cite{chomaz2025_review}, not all experiments necessarily require such extra contribution~\cite{petter_ferlaino_no_LHY} (see also Appendix F in~\cite{kirkbycomplexlangevin}), whose role in fact appears to decrease with increasing temperature~\cite{bisset_T_dipolar,kirkbycomplexlangevin}.
We defer all further discussion of this term, and its inclusion in phenomenological stochastic studies~\cite{bland_poli_22} to Sec.~VI and Appendix~B, with Eq.~(\ref{F-dipolar}) giving a plausible criterion for its relative importance in a finite-temperature dipolar gas.
\item (ii) The high-lying thermal, or `incoherent' modes obey a {\bf quantum Boltzmann equation} in the form
\begin{eqnarray}
\frac{\partial f}{\partial t} + \frac{\mathbf{p}}{m} \cdot\nabla f - \nabla \bigg(V_{ext}(x)
+ V_{nc}(x)\bigg) \cdot \nabla_{\mathbf{p}} f =
\frac{1}{2} (I_a + I_b) \label{zng-full} \;,
\end{eqnarray}
with the high-lying thermal modes of the system modelled through a distribution function $f(t,\mathbf{r},\mathbf{p})$ in phase space.
\item (iii) Non-local interactions are included through {\bf a Poisson-like equation} defining the effective dipolar potential $V_d$ in the form
\begin{eqnarray}
\frac{\nabla^2}{(n \cdot \nabla)^2} V_d (x) = \left( n_c (x) +
\tilde{n} (x) + \frac{1}{2} \xi_2 \right) - \int d^4 x' \Pi^R (x', x)
V_{nc} (x') \;. \label{poisson-full}
\end{eqnarray}
\end{itemize}

In the above equations, $g$ and $C_{dd}$ denote the usual local (s-wave) and non-local (dipolar) interaction strengths, $n_c$ and $\tilde{n}$ denote the condensate and non-condensate densities, $V_c$ and $V_{nc}$ the effective potentials seen by the condensate and the non-condensate, $I_a$, $I_b$ are appropriate collisional integrals -- respectively defined in Eqs.~\eqref{nc-def}, \eqref{ntilde-def}, \eqref{Vc-def}, \eqref{Vnc-def}, \eqref{eqcolls-a}, \eqref{eqcolls-b} -- and $V_{ext}$ is the  external potential.

The 3 lines in the dynamical classical field equation \eqref{sgpe-full-1}-\eqref{sgpe-full-3} for $\Phi_0$  amount respectively to a pure mean field dynamics term [\eqref{sgpe-full-1}], and two elastic collisional processes respectively, namely (i) an $O(g^2)$ dissipative contribution facilitating particle transfer {\em between} the coherent (classical field) modes and the incoherent (high-lying) modes described by the Boltzmann equation, with its corresponding noise term $\xi_1$ [\eqref{sgpe-full-2}] , and (ii) an $O(g^2)$ scattering process  between a coherent-band and an incoherent-band particle which however only leads to particle redistribution {\em within} each band and energy exchange {\em between} the bands, but not to direct particle transfer {\em between} those two bands, alongside its corresponding noise contribution [\eqref{sgpe-full-3}] .
The meaning of these terms is further discussed within the context of the stochastic projected GPE (SPGPE) model of Sec.~\ref{sec-spgpe}, a well-studied model where such terms first appeared, which emerges as a limiting case of our present theory\footnote{Note that the SPGPE model also makes explicit statements regarding the location of the cutoff relevant for its numerical implementations, with such aspects not directly addressed within our present work.}.

The number densities for the coherent and non-coherent part are
\begin{equation}
n_c=|\Phi_0|^2 \;, \label{nc-def}
\end{equation}
\begin{equation}
\quad \tilde{n}=\int \frac{d^3 p}{(2\pi)^3}f(x,\mathbf{p}) \,. \label{ntilde-def}
\end{equation}

The mean-field potentials for the coherent and incoherent parts are respectively
\begin{eqnarray}
V_c (x) &=& C_{dd} V_d (x) + \left( g -
\frac{C_{dd}}{3} \right) (n_c (x) + 2 \tilde{n} (x)) \;,  \label{Vc-def} \\
V_{nc} (x) &=& C_{dd} V_d (x) + 2 \left( g -
\frac{C_{dd}}{3} \right) (n_c (x) + \tilde{n} (x)) \;, \label{Vnc-def}
\end{eqnarray}
thus generalizing the usual mean-field potentials in the Hartree-Fock approximation by a term due to the long-range interaction.

The terms on the right-hand-side (RHS) of \eqref{zng-full} correspond to collisional terms of this Boltzmann equation and are given by
\begin{eqnarray}
I_a &=& 4 \left( g -
\frac{C_{dd}}{3} \right)^2 n_c \int \frac{d^3 p_1 d^3 p_2 d^3 p_3}{(2 \pi)^2}
\delta (\varepsilon_{c} + \varepsilon_{\mathbf{p}_1} -
\varepsilon_{\mathbf{p}_2} - \varepsilon_{\mathbf{p}_3}) \delta
(\mathbf{p}_2 - \mathbf{p}_1 - \mathbf{p}_c + \mathbf{p}_3)\nonumber\\
& & \hspace{2.5cm} \times (\delta (\mathbf{p}_1 - \mathbf{p}) - \delta (\mathbf{p}_2 -
\mathbf{p}) - \delta (\mathbf{p}_3 - \mathbf{p})) ((1 + f_1) f_2 f_3 -
f_1 (1 + f_2) (1 + f_3)) \label{eqcolls-a} \;, \\
I_b &=& 4 \left( g -
\frac{C_{dd}}{3} \right)^2 \int \frac{d^3 p_2 d^3 p_3 d^3 p_4}{(2 \pi)^5} \delta
(\varepsilon_{\mathbf{p}_3} + \varepsilon_{\mathbf{p}_4} -
\varepsilon_{\mathbf{p}_2} - \varepsilon_{\mathbf{p}}) \delta
(\mathbf{p} + \mathbf{p}_2 - \mathbf{p}_3 - \mathbf{p}_4) \nonumber\\
& & \hspace{2.5cm} \times [f_3 f_4 (f + 1) (f_2 + 1) - f f_2 (f_3 + 1) (f_4 + 1)] \label{eqcolls-b}\;.
\end{eqnarray}
The two functions $R$ and $\Pi^R$ in \eqref{sgpe-full-2}-\eqref{sgpe-full-3} are respectively
\begin{eqnarray}
R &=& \left( g -
\frac{C_{dd}}{3} \right)^2
\int \frac{d^3 p_1 d^3 p_2 d^3 p_3}{(2 \pi)^5} \delta
(\varepsilon_{\mathbf{q}} + \varepsilon_{\mathbf{p}_1} -
\varepsilon_{\mathbf{p}_2} - \varepsilon_{\mathbf{p}_3}) \delta (\mathbf{q} + \mathbf{p}_1
- \mathbf{p}_3 - \mathbf{p}_2)  \nonumber\\
& & \hspace{2.0cm} \times \bigg[ f_1 (1 + f_2) (1 + f_3) - (1 + f_1) f_2 f_3 \bigg]\nonumber\\
&=& \frac{1}{4 n_c} \int \frac{d^3 p}{(2 \pi)^3} I_a \label{rdef2} \;,
\end{eqnarray}
and, in terms of its Wigner transform, 
\begin{equation}
\Pi^R (x, \mathbf{k}) = \int \frac{d^3 p_1 d^3 p_2}{(2 \pi)^3}
\frac{1}{\varepsilon_{\mathbf{k}} + \varepsilon_{\mathbf{p}_2} -
\varepsilon_{\mathbf{p}_1} + i \sigma} \delta (\mathbf{k} + \mathbf{p}_2
- \mathbf{p}_1) \bigg[f_1 (1 + f_2) - f_2 (1 + f_1)\bigg]\,, \label{Pi-R}
\end{equation}
where the infinitesimal $i\sigma$ term indicates the way for the integration to be pefomed around the pole. The quantities $\xi_1$ and $\xi_2$ are Gaussian stochastic forces with correlation functions
\begin{eqnarray}
\langle \xi_1^{\ast} (x) \xi_1 (x') \rangle &=& \frac{i}{2} \Sigma_{(c)}^K (x)\delta(x-x')  \label{realcorrelators-a}  \;, \\
\langle \xi_2 (x) \xi_2 (x') \rangle&=&- 2 i \Pi^K (x, x') \;, \label{realcorrelators-b}
\end{eqnarray}
where the Wigner transforms of $\Sigma^K$ and $\Pi^K$ are 
\begin{eqnarray}
\Sigma_{(c)}^K(x) &=& - 2 i \left( g -
\frac{C_{dd}}{3} \right)^2
\int \frac{d^3 p_1 d^3 p_2 d^3 p_3}{(2 \pi)^5} \delta
(\varepsilon_{c} + \varepsilon_{\mathbf{p}_1} -
\varepsilon_{\mathbf{p}_2} - \varepsilon_{\mathbf{p}_3}) \delta (\mathbf{p}_c
+ \mathbf{p}_1 - \mathbf{p}_2 - \mathbf{p}_3) \nonumber\\
& & \hspace{3.0cm} \times \bigg[f_1 (1 + f_2) (1 + f_3) + (1 +
f_1) f_2 f_3\bigg] \label{sigkdef} \;,
\end{eqnarray}
and
\begin{eqnarray}
\Pi^K (x, \mathbf{k}) &=&  i \int \frac{d^3 p_1 d^3 p_2}{(2 \pi)^2} \delta
(\varepsilon_{\mathbf{k}} + \varepsilon_{\mathbf{p}_2} -
\varepsilon_{\mathbf{p}_1}) \delta (\mathbf{k} + \mathbf{p}_2 -
\mathbf{p}_1) \bigg[f_1 (1 + f_2) + f_2 (1 + f_1)\bigg] \label{pikdef} \;.
\end{eqnarray}

The above equations \eqref{sgpe-full-1}-\eqref{poisson-full}
are general equations that can reduce to certain known limits and they can be connected with other formalisms in the literature. For clarity,
%
we highlight once more that
as our finite temperature derivation based on~\cite{proukakis_rigopoulos_arxiv_24} considers fluctuations of the fast $\varphi$-modes only under the assumption of $(u_k=1\,, v_k=0)$, it does {\em not} self-consistently generate\footnote{We note that while the LHY term  can be routinely generated in a typical density-functional manner by adding a further term of the form $(2/5)\int d^3{\bf r} \gamma_{QF} |\Phi_0|^5$   to our energy functional~\cite{Aybar_2019}, i.e.~considering a more general effective action than that of the usual $\phi^4$ theory, we have not done this here as such a term is not {\em explicitly} additionally present in the original 
Hamiltonian, but its effect should emerge from the full Hamiltonian by appropriate handling (full Bogoliubov diagonalization).} 
the LHY correction term [Eq.~(\ref{extra})].
In the remainder of this paper, we first consider the well-studied case of typical ultracold quantum gases, interacting only via local effective interactions (s-wave scattering, Sec.~V), and then focus on the case of dipolar interactions (Sec.~VI), discussing common theories emerging as limiting cases.


\section{Reductions to Known Models for Locally-Interacting Atomic Gases \label{sec:contact}}

The local interaction limit is obtained by setting $C_{dd}=V_d=0$ in the above equations, for which we now discuss emerging common theories as limiting cases. 

\subsection{Mean Field Limit: The Gross-Pitaevskii equation}

In the simplest mean field limit we can consider that almost all bosons are in the condensed state, so we can neglect the effect of the incoherent particles. In this limit, we immediately recover the celebrated Gross-Pitaevskii equation
\begin{equation}
i \frac{\partial \Phi_0}{\partial t} = \left[ - \frac{1}{2 m} \nabla^2 +
V_{ext} + g\, |\Phi_0|^2 \right] \Phi_0 \;. \label{gpe}
\end{equation}

Starting from Eq.~\eqref{gpe}, we can also include excitations of the condensate via a linearised treatment around $\Phi_0$, leading to the well-known Bogoliubov-de Gennes equations~\cite{pitaevskii_1,pitaevskii_2,fetter_72,RevModPhys.71.463,proukakis_jackson_08,Pitaevskii,Pethick_Smith_2008}, containing quasiparticle physics.
In the dilute regime $n a^3 \ll 1$, such contributions are generally small. However, as we shall see later, this point will become very relevant in the subsequent discussion of Sec.~\ref{sec-dipolar} in the context of dipolar gases.

\subsection{The Zaremba-Nikuni-Griffin (or ZNG) Model: \\ A Self-Consistent Gross-Pitaevskii Boltzmann Description}

Accounting for incoherent (thermal) particles, but dropping all stochastic contributions\footnote{Formally, this is equivalent to working only up to order one in the quantum component of the Keldysh formalism.} and the $\Pi^R$ `scattering' terms we obtain 
\begin{eqnarray}
& & i \frac{\partial \Phi_0}{\partial t} = \left[ - \frac{1}{2 m} \nabla^2 +
V_{ext} + g (n_c + 2 \tilde{n}) - i R \right] \Phi_0 \label{zng-gpe}\\ 
& & \frac{\partial f}{\partial t} + \frac{\mathbf{p}}{m}\cdot \nabla f - \nabla \bigg(V_{ext} + 2 g (n_c + \tilde{n})\bigg) \nabla_{\mathbf{p}} f =
\frac{1}{2} (I_a + I_b) \label{zng-qbe}
\end{eqnarray}
where the dissipative contribution $R$ in the finite-temperature GPE, and the collisional integrals $I_a$ and $I_b$ in the quantum Boltzmann equation have the form given by Eqs.~\eqref{eqcolls-a}, \eqref{eqcolls-b}, \eqref{rdef2}, but with $C_{dd}=0$.
This set of equations corresponds to the equations of the so-called Zaremba-Nikuni-Griffin, or `ZNG' model~\cite{zaremba1999dynamics}, with extensive formulation and implementation details discussed in Ref.~\cite{griffin_nikuni_zaremba_2009}.

Such a set of equations, which makes an explicit separation into a well-formed condensate mode, with all remaining particles treated as thermal,
includes both the $O(g)$ self-consistent condensate-thermal dynamics (which already contains dissipative effects due to dynamical coupled mean field evolution), and the $O(g^2)$ collisional dynamics facilitating particle-transfer between the condensed and thermal sub-components. 
A brief review of such equations can be found in~\cite{ev2,proukakis_jackson_08} while the
 numerical scheme solving those equations has been discussed in~\cite{jackson_zaremba_02a,griffin_nikuni_book_09}.
 Importantly, in order to solve these equations, one has to first obtain a self-consistent equilibrium solution containing both condensed and thermal particles via the established Hartree-Fock model: in such a model, the static condensate spatial distribution $n_{c,0} = |\phi_0({\bf r})|^2$
 is obtained from a generalized static GPE for $\phi_0$ in the form 
\begin{equation}
\left[ - \frac{1}{2 m} \nabla^2 +
V_{ext} + g (n_{c,0} + 2 \tilde{n}_{0})  \right] \phi_0 = \mu_0 \phi_0  \label{gpe-static-HF} \;,
\end{equation}
while the thermal cloud density $\tilde{n}_0$ is obtained semi-classically in the local-density approximation via
\begin{equation}
    \tilde{n}_{0}({\bf r}) = \int \frac{d {\mathbf p}}{(2 \pi)^3} \,\,
    \frac{1}{e^{(\epsilon_p-\mu_0)/T}-1}
    \label{HF-thermal}
\end{equation}
where
\begin{equation}
   \epsilon_p = \frac{p^2}{2m} + V_{ext} + 2 g (n_c ({\bf r}) + \tilde{n} ({\bf r})) - \mu_0
   \label{HF-energies}
\end{equation}
is the effective Hartree-Fock energy for the thermal particles.
For further information on the iterative solution to these Hartree-Fock equations, performed at fixed temperature and total atom number, until reaching desired convergence on the chemical potential and atom number, see, e.g.~\cite{giorgini_HF_96,giorgini_HF_97,jackson_zaremba_02a,griffin_nikuni_book_09,allen_thesis_12}.

This generalised kinetic model has  been used to successfully model a range of dynamical condensate excitations, e.g.~\cite{ZNG4,ZNG-scissors,griffin_nikuni_book_09,Straatsma_2016}.
As this model is typically derived by relying on the concept of symmetry-breaking, and thus the existence of a well-formed largely-coherent proto-condensate, it can only describe dynamics outside the region of critical fluctuations. As such, it is not applicable to model phase-transition physics, or low-dimensional systems where phase fluctuations play a significant (dominant) role. Nonetheless, this model can produce useful (and seemingly accurate) results for condensate growth~\cite{ZNG-growth,ev1}, but only once an initial seed has been assumed in the numerical implementation (since the appearance of $\Phi_0$ as a common factor of all terms on the RHS of Eq.~\eqref{zng-gpe} implies that an initial $\Phi_0=0$ condition guarantees that $\Phi_0$ remains zero in all subsequent evolution, i.e.~no {\em ab initio} condensate growth can emerge, unless a proto-condensate seed is introduced into the model. Moreover, while the model is {\em not} valid in the pure 1D and 2D limits, it can be used as one approaches these regimes, provided the particle scattering is still of a kinematically-3D nature (and phase fluctuations are not dominant). In this context, it has also been successfully used to describe dissipation and reconnections of vortices in the quasi-2D limit~\cite{vortexT1,vortexT2,vortexT3}, and dark soliton dynamics in the quasi-1D limit~\cite{ZNG-soliton}.
For other implementations to experimental studies, see, e.g.~\cite{Xhani20,xhani_proukakis_22,ev2,lee_jorgensen_16,lee_jorgensen_18}. 

\subsection{Stochastic models}

The next class of well-studied equations that emerges as a limiting case of our full equations also contain stochastic noise terms. There are broadly two different models being discussed in the literature~\cite{stoof_99,blakie_bradley_08}: despite their different derivations, both these models rely on an effective separation into slow (`coherent') and fast (`incoherent') modes, a notion which is practically the same as that used in our own derivation~\cite{proukakis_rigopoulos_arxiv_24}. Although, in some limiting regime, both such equations lead to effectively the same final model, there are sufficient distinctions between the two approaches for us to briefly discuss such models separately below.

\subsubsection{Stochastic Approach of Stoof}

The stochastic approach of Stoof, developed in a series of papers~\cite{stoof_97,stoof_chapter_98,stoof_99} is based on a formulation of Keldysh non-equilibrium formalism in terms of coherent states,  where the main object of interest is the probability distribution functional and its Fokker-Planck equation.
The final set of equations is closely related to a subset of our general derived equations [\eqref{sgpe-full-1}-\eqref{poisson-full}] for a self-consistently coupled finite-temperature stochastic GPE, coupled to a quantum Boltzmann equation, which arises by ignoring $\Pi^R$ and $\Pi^K$ in them. 
Written in our notation,
the equations obtained by Stoof take the form\footnote{See, for example, the general Fokker-Planck equation for the low-lying modes found in Eq.~(231) of~\cite{stoof_99}, or equivalently Eq.~(32) in~\cite{Duine_2001} which generates Eq.~\eqref{sgpe-stoof-general}, and the corresponding quantum Boltzmann equation Eqs.~(227)  of~\cite{stoof_99} to which this is coupled].}
\begin{eqnarray}
& & i \frac{\partial \Phi_0 (x)}{\partial t} = \left( - \frac{1}{2 m} \nabla^2 +
V_{ext} + V_c \right) \Phi_0 (x) - i R \Phi_0 (x) + \xi_1  \label{sgpe-stoof-general}
\end{eqnarray}
for the `coherent' part of the system, dynamically coupled to
\begin{eqnarray}
\frac{\partial f}{\partial t} + \frac{\mathbf{p}}{m} \cdot\nabla f - \nabla \bigg(V_{ext}(x)
+ V_{nc}(x)\bigg) \cdot \nabla_{\mathbf{p}} f =
\frac{1}{2} (I_a + I_b) \label{qbe-stoof-general}
\end{eqnarray}
for the incoherent part of the system.\footnote{Note that the full theory of Stoof is formulated in terms of a momentum-dependent $T$-matrix, and writing the equations this way here, we also ignore the momentum dependence of the atomic interaction strength.}

In order to avoid the complexities of solving the stochastic equation self-consistently with a dynamical quantum Boltzmann equation, before its numerical implementation, a further  approximation of thermal equilibrium was made~\cite{Duine_2001}, ultimately reducing Eq.~\eqref{sgpe-stoof-general} to a more computationally amenable formula of a {\em single} stochastic GPE equation near thermal equilibrium. This was made possible by noting that, for a Bose-Einstein distribution,
\begin{equation}
f(\varepsilon_i,\mu) = \frac{1}{e^{\beta (\varepsilon_i - \mu)} - 1}
\end{equation}
where $\beta=1/k_B T$ and  $\epsilon_i$ corresponds to the energy of the particle $i$,
one can obtain a formal fluctuation-dissipation relation between the dissipative term $iR$ [Eq.~\eqref{rdef2}] and the corresponding Keldysh self-energy $\Sigma^K$ [Eq.~\eqref{sigkdef}].
This requires a further approximation of large occupation numbers, for which we can approximate
\begin{equation}
\frac{1}{1 + 2 f} \approx \frac{1}{2} \beta (\varepsilon_i - \mu), 
\end{equation}
this explicitly obtaining the fluctuation-dissipation relation
\begin{eqnarray}
i R = - \frac{\beta}{4} \Sigma_{(c)}^K (\varepsilon_q - \mu) \label{iReq} \;.
\end{eqnarray}
Noting that $\varepsilon_q$ corresponds to the energy of the coherent element involved in the collisional process with a non-coherent particle and that the terms $R$ and $\Sigma_{(c)}^K$ can be written as local terms, we can write equation~\eqref{iReq} in an operator form as
\begin{equation}
i R \Phi_0 = - \frac{\beta}{4} \Sigma_{(c)}^K \bigg(- \frac{1}{2 m} \nabla^2 + V_{ext}
+ g (n_c + 2 \tilde{n}) - \mu \bigg)\Phi_0 \;.
\end{equation}
Using these relations in our equation \eqref{sgpe-stoof-general} and making the redefinitions $\Phi_0\to \Phi_0 e^{-i \mu t}$ and $\xi_1\to \xi_1 e^{-i \mu t}$, after some arrangements we get that~\cite{stoof_99,Duine_2001}
\begin{eqnarray}
\label{stochasticineq-0}
i \frac{\partial \Phi_0 (x)}{\partial t} &=& \left( 1 + \frac{\beta}{4}
\Sigma_{(c)}^K \right) \left[ - \frac{1}{2 m} \nabla^2 + V_{ext} (x) +
g (n_c (x) + 2 \tilde{n} (x)) - \mu \right] \Phi_0 (x) + \xi_1 (x) \;,
\end{eqnarray}
with the stochastic noise correlations
\begin{eqnarray}
\langle \xi_1^{\ast} (x) \xi_1 (x') \rangle = \frac{i}{2} \Sigma_{(c)}^K (x)\delta(x-x') \;. \label{stochastic-corr}
\end{eqnarray}
Eqs.~\eqref{stochasticineq-0}-\eqref{stochastic-corr} practically correspond to the stochastic Gross-Pitaevskii equation obtained by Stoof~\cite{stoof2001dynamics,Duine_2001}, except that here -- unlike the typical formalism of Stoof --
we have additionally included the mean-field term $2g \tilde{n}$. 
It is important to note that explicit inclusion of such a mean-field term in the SGPE is essential in order to obtain a quantitative {\em ab initio} description of the system and compare to experimental observations, as discussed, e.g.~in~\cite{Cockburn12Ab,Ota18Collisionless} (see also~\cite{Davis_2012}).

Relabelling the factor on the front of the RHS of Eq.~\eqref{stochasticineq-0} as
\begin{equation}
    \frac{\beta}{4} 
    \Sigma_{(c)}^K(x,t) \,\, \rightarrow \,\, - i \gamma(x,t) \;,
\end{equation}
dropping the incoherent mean field contribution $2 g \tilde{n}$, and 
additionally suppressing the position and time dependence of the dissipative factor $\gamma$ (consistent with numerical solutions to date which typically treat this as a constant), we obtain the more familiar equation
\begin{eqnarray}
i \frac{\partial \Phi_0 (x)}{\partial t} = \left( 1 -i \gamma \right) \left( - \frac{1}{2 m} \nabla^2 +
V_{ext} + g |\Phi_0|^2 - \mu   \right) \Phi_0 (x) + \xi_1  \label{sgpe-stoof-simple} \;,
\end{eqnarray}
which is usually termed the Stochastic Gross-Pitaevskii equation (SGPE).

Before commenting further on the use of such an equation in Sec.~\ref{sec-sgpe-predictions}, we note here that an equation like~\eqref{sgpe-stoof-simple} was also obtained as the limiting case of an alternative approach, briefly discussed below.


\subsubsection{Stochastic Approach of Gardiner {\em et al.} \label{sec-spgpe}}

An alternative and completely independent stochastic approach~\cite{Gardiner_2002a,Gardiner_2002b,Gardiner03The,blakie_bradley_08} was developed, around the same time, from the Gardiner-Zoller quantum kinetic theory~\cite{jaksch_gardiner_97,gardiner_zoller_98,gardiner2017quantum} based on techniques from quantum optics~\cite{gardiner2017quantum}.
While, on first inspection, this might seem rather distinct to both our method and that of Stoof, there are a number of key common considerations: most notably these are the separation into a band of low-lying modes containing the condensate and other highly-populated modes whose energies are affected by its presence -- which they termed the `condensate band', or the `classical region' -- and a band of higher-lying purely thermal modes, which are typically then also approximated as being at equilibrium. 
Moreover, this model\footnote{A partial generalization of this model including a self-consistently coupled quantum Boltzmann equation in the limit $\Pi^R=\Pi^K=0$ has already been discussed by Garrett and Proukakis (unpublished).} explicitly includes a projector ensuring  these two `bands' of modes remain orthogonal to each other, important for their  numerical implementation.
Compared to Stoof's treatment, their analysis includes an additional
collisional term, associated with an extra noise contribution, but no explicit handling of a quantum Boltzmann equation.

Repeating within our treatment the same near-equilibrium analysis discussed above, but now also explicitly maintaining the $\Pi^R$ and $\Pi^K$ terms of Eqs.~\eqref{Pi-R}, \eqref{pikdef},
we note that substitution of an equilibrium Bose-Einstein distribution
\begin{equation}
f(\varepsilon_i,\mu) = \frac{1}{e^{\beta (\varepsilon_i - \mu)} - 1}
\end{equation}
in the equations  for $\Sigma^K$ [Eq.~\eqref{sigkdef}] and $\Pi^K$ [Eq.~\eqref{pikdef}] yields the generalised equilibrium relations

\begin{eqnarray}
i R = - \frac{1}{2} \Sigma_{(c)}^K (x) \frac{1}{1 + 2 f (\varepsilon_q,\mu)} \hspace{1.0cm} {\rm and} \hspace{1.0cm}
\Pi^R(x, \mathbf{k})-\Pi^A(x, \mathbf{k}) = \Pi^K (x, \mathbf{k}) \frac{1}{1 + 2f(\varepsilon_k,0)} \;,
\end{eqnarray}
which, taking again the limit of large occupation number via
\begin{equation}
\frac{1}{1 + 2 f} \approx \frac{1}{2} \beta (\varepsilon_i - \mu), \hspace{1.0cm} {\rm and} \hspace{1.0cm} \frac{1}{1 + \frac{2}{e^{\beta \varepsilon_k} - 1}} \approx \frac{1}{2} \beta \varepsilon_k \;,
\end{equation}
yields the relations
\begin{eqnarray}
i R = - \frac{\beta}{4} \Sigma_{(c)}^K (\varepsilon_q - \mu) \hspace{1.0cm} {\rm and} \hspace{1.0cm}
\Pi^R(x, \mathbf{k})-\Pi^A(x, \mathbf{k}) = \frac{\beta}{2} \Pi^K (x, \mathbf{k}) \varepsilon_k  \;. \label{fd}
\end{eqnarray}

We have already shown above how the fact that $\varepsilon_q$ corresponds to the energy of the coherent element involved in the collisional process with a non-coherent particle leads to such an equation for $iR$.
For $\Pi^R$ the situation is slightly different. Using that 
\begin{equation}
\Re (\Pi^A (x, \mathbf{k})) = \Re (\Pi^R (x,
\mathbf{k})) \hspace{1.0cm} {\rm and} \hspace{1.0cm} \Im (\Pi^A (x, \mathbf{k})) = - \Im (\Pi^R
(x, \mathbf{k}))
\end{equation}
we can recast the second equation of \eqref{fd} into 
\begin{equation}
    - \frac{\beta}{4} i \Pi^K (x, \mathbf{k}) \varepsilon_k = \Im (\Pi^R
(x, \mathbf{k})) \;.
\end{equation}
Thus, the term containing $\Pi^R$ can be split into a real and imaginary part,
\begin{equation}
    \Pi^R(x,\mathbf{k}) = 
           \Re (\Pi^R (x,\mathbf{k}))
        +i \, \Im (\Pi^R (x, \mathbf{k})) \;.
\end{equation}
We will focus only on the imaginary part, which relates to $\Pi^K$, and gives the known scattering part. This can be understood upon noting that $\varepsilon_k$ is an energy with a value within a coherent band, which does not correspond to a single particle energy, but rather a combination of slow energies. 
We interpret the process of this piece corresponding to an interaction between a coherent element with an incoherent one, changing their energies without changing their number, which implies that the energy $\varepsilon_k$ must be the difference between the out and in energies of the coherent element. Therefore, we can write
\begin{equation}
\varepsilon_k = \varepsilon_{q_1} - \varepsilon_{q_2} = \left( \frac{q^2_1}{2
m} + V_c \right) - \left( \frac{q^2_2}{2 m} + V_c \right) = \frac{q^2_1}{2 m}
- \frac{q^2_2}{2 m} \;.
\end{equation}
Since the operator $\Pi^K$ will be acting on $n_c(x')=|\Phi_0|^2$, the momenta $q_1$ and $q_2$ correspond to the momenta of the $\Phi^\ast(x')$ (out) and $\Phi(x')$ (in). With all of this in mind, we can write the $\Pi^R$ relation of Eq.~\eqref{fd} in an operator way as
\begin{equation}
\Im(\Pi^R(x',x) )\,n_c(x')\, \Phi_0 (x)= -\frac{\beta}{4} \, i \Pi^K (x', x) \left( \frac{1}{2 m} \Phi_0(x')
\nabla^2 \Phi_0^{\ast}(x') - \frac{1}{2 m} \Phi_0^{\ast}(x') \nabla^2 \Phi_0(x') \right)\Phi_0(x) \;.
\end{equation} 
Using these relations in our equation \eqref{sgpe-full-1}-\eqref{sgpe-full-3}, ignoring the term $2g\tilde{n}$, and making, as before, the redefinitions $\Phi_0\to \Phi_0 e^{-i \mu t}$ and $\xi_1\to \xi_1 e^{-i \mu t}$, after some arrangements we get that
\begin{eqnarray}
\label{stochasticineq}
i \frac{\partial \Phi_0 (x)}{\partial t} &=& \left( 1 + \frac{\beta}{4}
\Sigma_{(c)}^K \right) \left[ - \frac{1}{2 m} \nabla^2 + V_{ext} (x) +
g n_c (x) - \mu \right] \Phi_0 (x) + \xi_1 (x) \nonumber\\
& & +\beta g^2 
\int d^4 x' i\Pi^K(x,x') \nabla \cdot J (x')
\Phi_0 (x) + g \xi_2 (x) \Phi_0 (x) \;,
\end{eqnarray}
where we have defined
\begin{equation}
\nabla \cdot J (x') = \frac{i}{2 m} \Phi_0 (x') \nabla^2 \Phi_0^{\ast} (x')
- \frac{i}{2 m} \Phi_0^{\ast} (x') \nabla^2 \Phi_0 (x') \;.
\end{equation}
In relation to the reduced stochastic equation of Stoof [Eq.~\eqref{stochasticineq-0}], we observe the emergence of another kernel associated with an additional real multiplicative noise $\xi_2$.

The stochastic projected Gross-Pitaevskii equation (SPGPE) is usually expressed in the form
\begin{eqnarray}
\label{spgpeeq}
d \psi &=&\mathcal{P} [- i \mathcal{L} \psi d t] \nonumber \\ &+& \mathcal{P} [\beta G (\mathbf{r})
(\mu - \mathcal{L}) \psi d t + d W_G] \nonumber \\ &+& \mathcal{P} [- i V_M \psi d t + i \psi d W_M] \;, \label{spgpe-form}
\end{eqnarray}
where here $\mathcal{L}$ is given by
\begin{equation}
\mathcal{L} = - \frac{1}{2 m} \nabla^2 + V_{ext} + g n_c \;,
\end{equation}
and $\mathcal{P}$ is a projector that restricts the evolution of $\psi$ to the coherent region. 
%
The (full) SPGPE includes 
two noise terms $dW_G$ and $dW_M$ 
which  satisfy the relations
\begin{eqnarray}
\langle d W_G^{\ast} (x) d W_G (x') \rangle &=& 2 G (\mathbf{r}) \delta (x -
x') d t \;, \\
\langle d W_M (\mathbf{r}) d W_M (\mathbf{r}') \rangle &=& 2 M (\mathbf{r} - \mathbf{r}') d t \;.
\end{eqnarray}
For more details, see, e.g.,Ref.~\cite{rooney_collective_12} where full expressions for $G(\mathbf{r})$, $M(\mathbf{r})$ and $V_M(\mathbf{r})$ can also be found.

In the SPGPE formalism, the contribution associated with $G(r)$ is known as a growth term (consistent with Stoof's earlier discussion), and corresponds to an energy- and momentum-conserving binary collision which leads to a change in the number of particles within the $\Phi_0$ band.
The second term, associated with $M$, corresponds to a collision between a coherent band and an incoherent band particle, which shifts the energies, but does not lead to net population transfer between the two bands. 
As such, these terms -- whose explicit expressions can be found in~\cite{blakie_bradley_08} -- have been respectively referred to as `number-damping' and `energy-damping' contributions~\cite{rooney_collective_12,rooney_allen_16,bradley_spgpe_scipost_20,mehdi_hope_23}. Their relative importance is expected to depend on the particular physical scenario being considered, e.g.~energy damping (and its associated noise) typically dominate for near-equilibrium systems, except at low phase-space density~\cite{krause_24}.

To demonstrate the correspondence of our reduced equation~\eqref{stochasticineq} with the above form of the SPGPE, we first re-arrange the SPGPE, Eq.~\eqref{spgpe-form}, and make the redefinitions $\psi \to \psi e^{- i \mu t}$ and $d W_G \to d W_G \, e^{- i \mu
t}$, to recast the SPGPE in the equivalent form
\begin{equation}
\label{spgpeeq2}
i d \psi =\mathcal{P} \left[\left(1 - i \beta G (\mathbf{r})\right) (\mathcal{L} - \mu) \psi d t + i
d W_G +V_M \, \psi dt - \psi \, d W_M \right] \;.
\end{equation}

Using now the relations
\begin{equation}
G = \frac{i}{4} \Sigma_{(c)}^K (x) \, , \hspace{1.0cm} {\rm and} \hspace{1.0cm}  M = - i g^2 \Pi^K \;,
\end{equation}
and under the further identification 
\begin{equation}
i d W_G = \xi_1  \, , \hspace{1.5cm} {\rm and} \hspace{1.5cm}  - d W_M = g \xi_2 \;,
\end{equation}
it is easy to see that both formalisms become equivalent, upon also dropping the explicit appearance of the projector of the SPGPE formalism.
Note that  as both ourselves and Stoof are working  in a path integral formalism, such projectors do not appear explicitly, but the existence of these two distinct subspaces is implicit from the outset within the path integral approach, a feature becoming relevant in numerical implementations\footnote{For example, explicit inclusion of a projector sets a natural band limit to the noise and is important in any such numerical implementations, which will be addressed in future work. Note that the damping parameters depend on the cutoff, which also feeds into the reservoir parameters, with the cutoff usually set at around $2-3 \mu$.}.

Ignoring the so-called scattering term effectively reproduces Eq.~\eqref{sgpe-stoof-simple}, but now with the projector, $\mathcal{P}$, appearing explicitly in the equation; further redefining $\psi \to \psi e^{ i \mu t}$ and $d W_G \to d W_G \, e^{ i \mu
t}$, thus leads to the so-called `simple growth' SPGPE in the form
\begin{eqnarray}
\label{spgpeeq3}
d \psi &=&\mathcal{P} [- i \mathcal{L} \psi d t] +\mathcal{P} [\beta G (\mathbf{r})
(\mu - \mathcal{L}) \psi d t + d W_G]  \;.
\end{eqnarray}
%
This equation is thus practically equivalent to the earlier form discussed  as a limiting case of Stoof's theory [Eq.~\eqref{sgpe-stoof-simple}].

For completeness, we also note here in passing that an equation 
effectively the same as the reduced Stoof equation [Eq.~\eqref{sgpe-stoof-simple}] or the `simple growth' S(P)GPE [Eq.~\eqref{spgpeeq3}] can be alternatively
 referred to as a stochastic time-dependent Ginzburg-Landau theory. 
See, for example, the discussion of its time-independent version in Ref.~\cite{berloff_brachet_14} (and references therein), where a slightly different implementation of the projector is highlighted.

\subsubsection{Stochastic (Projected) Gross-Pitaevskii Predictions \label{sec-sgpe-predictions}}

The stochastic Gross-Pitaevskii equation (both with and without an explicit projector in the numerics) has been used to study condensate growth dynamics~\cite{stoof2001dynamics}, atom laser operation~\cite{Proukakis03Coherence,lee_haine_15}, rotating condensates~\cite{Bradley08Bose}, spontaneous defect generation and Kibble-Zurek physics~\cite{Weiler08Spontaneous,Proukakis09The,Damski10Soliton,Su13Kibble,Rooney13Persistent,Kobayashi16Thermal,Kobayashi16Quench,Liu18Dynamical,Liu20Kibble,bland_marolleau_20}, dynamical post-quench relaxation~\cite{Comaron19Quench,groszek_comaron_PRR,brown_bland_21}, 
and dynamics of
soliton~\cite{cockburn_nistazakis_10,cockburn_nistazakis_11,spgpe_soliton_2011}, vortex~\cite{rooney_bradley_10,rooney_allen_16,mehdi_hope_23}, persistent current~\cite{Rooney13Persistent,bland_marolleau_20,mehdi_bradley_21}, 
sound propagation~\cite{Ota18Collisionless},
and collective modes~\cite{rooney_collective_12,Straatsma_2016,bradley_spgpe_scipost_20}.
In addition to these there is a
plethora of equilibrium studies~\cite{Stoof02Low,proukakis_06b,cockburn_gallucci_11,cockburn_negretti_11,gallucci_cockburn_2012,Cockburn12Ab,Garrett_2013,Henkel_2017}, and mixture/spinor dynamics~\cite{bradley_blakie_14,liu_stoch_mixtures_16,Roy_2021,Roy_2023}, with related implementations discussed in~\cite{proukakis_schmiedmayer_06,swislocki_deuar_2016,deuar_2016,deuar_2017,deuar_2019,thomas_davis_relaxation,bradley_low_D_spgpe,keepfer_liu_22}.

As touched upon earlier, explicit analytical predictions exist for the various growth and scattering terms in both methods, with detailed expressions reviewed  in~\cite{Duine_2001,blakie_bradley_08}.
However, despite some very interesting claims~\cite{Rooney13Persistent,bradley_spgpe_scipost_20}, it is not yet universally accepted whether the analytically-computed expressions can model (all) experiments {\em quantitatively}.
Combined with the fact that current numerical treatments do not yet include the (full) dynamics of the incoherent band, a common approach in the literature is to replace $\gamma$ by a constant in both position and time, with a typical value (well) within the (analytically-estimated) range $[10^{-4},\,10^{-1}]$.
This is mostly the case with equations ignoring the scattering terms.
In that case, the stochastic simulations can be ``normalized" to a given experiment by identifying a chosen probed observable (usually the condensate number growth~\cite{Weiler08Spontaneous,Liu18Dynamical}), and choosing a {\em constant} (so both position- and time-independent) value of $\gamma$ such that the effective numerical dynamics reproduces the experimental observations to the desired accuracy: such $\gamma$ value is then typically kept fixed, with temperature recorded in the amplitude of the related noise fluctuations. However, it should also be noted here that SPGPE approaches which include both growth and scattering contributions typically use analytically-calculated values for such parameters. While this is of course preferrable, some questions remain regarding the extent to which this can give fully {\em quantitative} predictions.

\subsubsection{The Dissipative (or Phenomenologically Damped) Gross-Pitaevskii Equation}

The simplicity of Eq.~\eqref{sgpe-stoof-simple} above makes it tempting to work simply with a {\em phenomenologically} damped GPE (DGPE), which can be directly obtained from this, upon ignoring the additive noise contribution. In that limit, we obtain the form
\begin{eqnarray}
i \frac{\partial \Phi_0 (x)}{\partial t} = \left( 1 -i \gamma \right) \left( - \frac{1}{2 m} \nabla^2 +
V_{ext} + g |\Phi_0|^2 - \mu \right) \Phi_0 (x)  \;.
\end{eqnarray}
Remarkably, such an equation was first proposed, on purely phenomenological grounds, by Pitaevskii~\cite{pitaevskii_3,pitaevskii_4}. It was re-introduced into the cold-atom community in Refs.~\cite{choi_morgan_98,tsubota_kasamatsu_03} and has become the {\em toy} model for thermal dissipation, providing in many cases very good, albeit {\em qualitative} dynamics. A comparison of the (single) DGPE and the average of the S(P)GPE trajectories has been performed in the context of dissipative dark soliton~\cite{cockburn_nistazakis_10} and vortex~\cite{rooney_allen_16} dynamics, showcasing also the dominant ``destabilizing" effects of the noise. The significant benefit of the simplified DGPE model is that it can be very easily applied numerically, and as it is engineered to (ultimately) yield the correct $T=0$ equilibrium system for a given atomic mass, trap configuration, interaction strength, and chemical potential, it can thus generally model (in a phenomenological manner) the dissipation of excitations (e.g. sound, solitons, vortices) above the ground state.

Having shown how the various models typically used in the ultracold atomic community for modelling weakly-interacting gases interacting via contact potential emerge naturally as limiting cases of our generalised model of Eqs.~\eqref{sgpe-full-1}-\eqref{poisson-full}, we now proceed to discuss the more general context of finite-temperature dipolar gases, for which the finite-temperature modelling is not as advanced, and where our present work can offer new model equations for studying non-equilibrium processes in such systems.



\section{Corresponding Finite-Temperature Equations for Dipolar Gases \label{sec-dipolar} }

In the previous section, we have explicitly shown how our generalised stochastic formalism provides an appealing, more transparent, unifying way, to obtain a broad range of established theories for weakly-interacting atomic gases with contact interactions as limiting cases, based on different approximations.
Most notably, our formalism can be seen as creating a unifying framework  {\em simultaneously} incorporating {\em both} stochastic growth {\em and} scattering processes on the one hand, {\em and} quantum Boltzmann thermal particle dynamics on the other hand, in a dynamically self-consistent manner.
%
In this section we demonstrate the full power of our methodology, by extending our model to systems interacting with long-range dipolar interactions.

\subsection{The Gross-Pitaevskii Equation}

 The simplest approximation in the literature is, as before, to only consider pure mean field dynamics, but now in the context of the combined (local plus non-local) interaction potential. The Gross-Pitaevskii equation then takes the general form
\begin{equation}
i \frac{\partial \Phi_0}{\partial t} = \left[ - \frac{1}{2 m} \nabla^2 +
V_{ext} + \int d{\bf r'} U({\bf r}-{\bf r'}) |\Phi_0({\bf r'})|^2 \right] \Phi_0 \;,
\end{equation}
with the general interaction potential $U({\bf r}-{\bf r'})$ of Eq.~\eqref{U-s-nl}-\eqref{U-dd}, including both s-wave and dipolar contributions.

This equation describes the mean-field regime of dipolar gases in the near-$T=0$ limit. Recently, significant attention has been placed in modelling so-called dipolar droplets~\cite{ferrier_droplets,chomaz_droplets}, i.e.~self-bound `mini-condensates'  whose collapse is stabilised by quantum fluctuations through a quantum correction term to the mean-field energy, known as the Lee-Huang-Yang, or LHY, contribution~\cite{lee_huang_57,lima_pelster_11,lima_pelster_12,fischer_LHY,wachtler_santos}. 
Such beyond-mean-field physics is clearly not contained within the above GPE.
Nonetheless, there exists a generalized version of it, known as the extended GPE, or eGPE, which has quickly become the norm in this community~\cite{lima_pelster_11,lima_pelster_12,fischer_LHY,wachtler_santos,baillie_droplet,Bisset_2016,Chomaz_2022}, and which we therefore briefly discuss below.

\subsubsection{The Extended Gross-Pitaevskii Equation (eGPE) for Dipolar Condensates \label{sec-egpe}}

Following here directly the work of Lima and Pelster~\cite{lima_pelster_11,lima_pelster_12}) (see Appendix C for details), we
linearize the condensate wavefunction, following the usual Bogoliubov-de Gennes analysis, via
\begin{equation}
\Phi (r, t) = 
 e^{- i \mu t} \left( \Phi_0 (r) + \eta(r,t) \right)
= e^{- i \mu t} \left( \Phi_0 (r) +  \underset{k}{\sum} \left(u_k (r) \chi_k (t) + v^{\ast}_k (r) \chi^{\ast}_k(t) \right) \right) \;.
\end{equation}
This gives rise to a rather simple-looking equation which incorporates quantum fluctuations as a simple additional contribution into the standard mean-field GPE, which now takes the form
\begin{equation}
i \frac{\partial \Phi_0}{\partial t} = \left[ - \frac{1}{2 m} \nabla^2 +
V_{ext} + \int d{\bf r'} U({\bf r}-{\bf r'}) |\Phi_0({\bf r'})|^2 \right] \Phi_0 + \gamma_{QF} |\Phi_0|^3 \Phi_0 \;. \label{egpe} 
\end{equation}
The latter contribution
incorporates the LHY quantum fluctuation contribution in terms of the diluteness parameter $(na^3)$.
In the simplest, zero-temperature, limit, this takes the form
\begin{equation}
\gamma_{QF} = \frac{32}{3} g \sqrt{\frac{a^3}{\pi}} \mathcal{Q}_5 (\epsilon_{dd}) \;, \label{gamma-qf}
\end{equation}
Equation~\eqref{egpe} is known as the {\em extended} Gross-Pitaevskii equations (eGPE), and has become the norm for studying beyond-mean-field dipolar condensates at near-zero temperatures~\cite{Chomaz_2022}.
In the above expressions,
\begin{equation}
	\mathcal{Q}_5 (\epsilon_{dd}) = \int^1_0 d u (1 - \epsilon_{dd}
	+ 3 \epsilon_{dd} u^2)^{5 / 2} \;,
\end{equation}
and $\epsilon_{dd}=C_{dd}/3g$ is a dimensionless parameter relating the two interaction strengths $C_{dd}$ and $g$ also used in the literature.

Further modifications to the prefactor $\mathcal{Q}_5 (\epsilon_{dd})$ are imposed by the presence of a (static) thermal cloud~\cite{Boudjemaa_2016,Boudjemaa_2017,Aybar_2019} and special care is needed to correctly identify the precise location of the momentum-space cutoff in evaluating relevant expressions~\cite{maucher_dipolar}. While such terms should additionally be considered, we note that the effect of temperature is simply to replace the factor $\mathcal{Q}_5 (\epsilon_{dd}) $ by a more general, temperature-dependent, factor, featuring an {\em additional} contribution scaling with $T^2$~\cite{Boudjemaa_2016,Boudjemaa_2017,Aybar_2019,maucher_dipolar}. 
Thus, while acknowledging the existence of such further terms, for the sake of our arguments we shall limit our discussion of quantum fluctuations here to the zero-temperature correction prefactors  $\mathcal{Q}_5 (\epsilon_{dd}) $, bearing such amendment in mind, which can easily be implemented at a later stage if deemed necessary.

This section has outlined an established proof of how the equation for the pure condensate mean-field can be systematically extended to account for the role of {\em quantum} fluctuations, in the regime where such terms become important. However, it is pertinent to remind the reader that all other formal considerations within this work have focussed on the rather distinct regime where {\em thermal} fluctuations dominate. It is therefore plausible to expect that such two distinct pictures, valid in separate limits of the relative ratio of importance of thermal to quantum effects, can be somehow combined, and we comment on this further below (Sec.~\ref{eQBE}).

\subsection{Gross-Pitaevskii-Boltzmann, or ZNG Model of Dipolar Condensates \label{sec:dipolar-zng}}

Returning to our self-consistently-derived formalism now [Eqs.~\eqref{sgpe-full-1}-\eqref{poisson-full}, without the added LHY corrections], and ignoring initially the scattering $\Pi^R$ terms and all stochastic noises, our full set of self-consistently obtained equations reduces to a dissipative GPE and a quantum Boltzmann set of equations, self-consistently coupled to a Poisson-like equation parametrizing the long-range interactions, in the form\footnote{Note that, for enhanced clarity, we have made the appearance of $V_d$ explicit in these equations.}
\begin{eqnarray}
& & i \frac{\partial \Phi_0 (x)}{\partial t} = \left( - \frac{1}{2 m} \nabla^2 +
V_{ext}(x) +   C_{dd} V_d (x) + \left( g -
\frac{C_{dd}}{3} \right) (n_c (x) + 2 \tilde{n} (x))  \right) \Phi_0 (x) 
- i R \Phi_0(x) \label{full-dipolar-zng-gpe} \\ 
& & \frac{\partial f}{\partial t} + \frac{\mathbf{p}}{m} \cdot\nabla f - \nabla \bigg(V_{ext}(x) +  C_{dd} V_d (x) + 2 \left( g -
\frac{C_{dd}}{3} \right) (n_c (x) + \tilde{n} (x))\bigg) \cdot \nabla_{\mathbf{p}} f = \frac{\left( I_a + I_b \right)}{2} \label{full-dipolar-zng-qbe} \\
& & \nabla^2 V_d (x) = (\mathbf{n} \cdot \nabla)^2\bigg(n_c (x) + \tilde{n} (x) \bigg) \label{full-dipolar-zng-Vd} \;.
\end{eqnarray}
Such a system of equations can be thought of as the ZNG kinetic model for dipolar condensates, and is expected to provide an accurate description of non-equilibrium thermal effects in dipolar condensates when the quantum fluctuations are sub-dominant.

We stress that the equations presented in this section arise directly from a formalism explicitly focussing on thermal fluctuations: in the regime where such thermal fluctuations are important, leading to corrections on the otherwise pure-condensate physics, it might be reasonable to expect that quantum fluctuations will typically be sub-dominant. As such, it might be that the most relevant regimes in ultracold dipolar systems are either those with condensate and thermal fluctuations (such as the dipolar ZNG equations of Eqs.~\eqref{full-dipolar-zng-gpe}-\eqref{full-dipolar-zng-Vd}), or those with condensate and quantum fluctuations (such as the extended GPE of Eq.~\eqref{egpe}), essentially choosing between the generalised finite-temperature models self-consistently derived in an {\em ab initio} manner in our work, or,  at lower temperatures, to the established LHY quantum correction term summarized in Sec.~\ref{sec-egpe} and leading to the extended GPE of Eq.~\eqref{egpe}.

While this is still a rather involved set of equations to be solved numerically, one might -- for example -- in first instance be interested in studying the damping of collective modes in a regime where mutual friction between the condensate and thermal cloud dominates, while particle exchange, or scattering, between these two sub-components is not significant\footnote{A somewhat related setting of finite-temperature dynamical equations in the so-called Hartree-Fock-Bogoliubov limit, which explicitly includes mean-field dynamics and anomalous contributions but does not include collisional redistribution, has been discussed in~\cite{Boudjemaa_2020}.}: in that case, one could potentially further ignore all second-order interaction collisional terms.
Setting $iR = I_a = I_b = 0$ in the above equations would correspond to the collisionless dipolar ZNG discussed below.

\subsection{An Extended Gross-Pitaevskii, Collisionless Boltzmann Equation Model? \label{eQBE}}

Considering the {\em somewhat} `orthogonal' directions of travel presented in Sec.~\ref{sec-egpe} (quantum fluctuations) and Sec.~\ref{sec:dipolar-zng} (thermal fluctuations) -- each expected to be valid in the relevant regime --, and noting that the latter approach on which our main equations are based does {\em not} include condensate fluctuations to all orders, one might consider postulating a combination of those two pictures, albeit in a somewhat {\em ad hoc} manner, to interpolate between such regimes.

What we envisage here is {\em supplementing} our above dipolar ZNG equations for finite-temperature dipolar condensates (but, at least in first instance, in the collisionless limit $iR = I_a = I_b = 0$) by the explicit {\em addition} of a term of the form $\gamma_{QF} |\Phi_0|^3 \Phi_0$ in the equation describing the {\em condensate} mean field, i.e.~in Eq.~\eqref{full-dipolar-zng-gpe} (without the $-iR \Phi_0$ term). 
Such an approach would be broadly in line with the most advanced current approaches for finite temperature dipolar gases with explicit quantum contributions. In that context we note static $T>0$ approaches based on self-consistent Hartree-Fock-Bogoliubov-type treatments~\cite{Boudjemaa_2016,Boudjemaa_2017,Aybar_2019,maucher_dipolar} (building upon ~\cite{Ronen_2007}) or a treatment introducing an additional finite-temperature contribution to the grand canonical potential~\cite{politi_pohl_23}, both of which lead to a finite-temperature eGPE which includes an additive thermal contribution term, appearing in the eGPE in a similar manner to the quantum LHY contribution. Interestingly, such an approach has been taken further in the dynamical realm~\cite{bland_poli_22} (building upon~\cite{linscott_blakie_14}), by utilizing -- in a somewhat {\em ad hoc} manner (see also discussion below) -- a stochastic equation having the form of the previously-discussed (`simple-growth') S(P)GPE [Eq.~(\ref{sgpe-stoof-simple}), or equivalently, (\ref{spgpeeq})] but additionally explicitly including ({\em heuristically}) the standard LHY correction. Remarkably, such an extended (postulated) equation appears to give very good agreement with experiments~\cite{bland_poli_22}, indicating that it might in fact be reasonable (even if not necessarily exact) to combine both quantum (eGPE) and thermal (SPGPE) approaches in such a manner.

While our formalism cannot offer any concrete proof supporting, or contradicting, such a 
procedure of simultaneously introducing both thermal and quantum fluctuations in a dynamical manner,
we do nonetheless feel that it would be beneficial for a careful future examination of the implications of such a postulated treatment against experimental evidence to be carried out.

With the above caveats, one might be tempted to {\bf {\em postulate} }the following set of extended collisionless ZNG equations
\begin{eqnarray}
& & i \frac{\partial \Phi_0 (x)}{\partial t} = \left( - \frac{1}{2 m} \nabla^2 +
V_{ext}(x) +   C_{dd} V_d (x) + \left( g -
\frac{C_{dd}}{3} \right) (n_c (x) + 2 \tilde{n} (x))  \right) \Phi_0 (x) +\gamma_{QF} |\Phi_0|^3\Phi_0 \label{dipolar-zng-egpe} \\ 
& & \frac{\partial f}{\partial t} + \frac{\mathbf{p}}{m} \cdot\nabla f - \nabla \bigg(V_{ext}(x) +  C_{dd} V_d (x) + 2 \left( g -
\frac{C_{dd}}{3} \right) (n_c (x) + \tilde{n} (x))\bigg) \cdot \nabla_{\mathbf{p}} f = 0 \label{dipolar-zng-qbe} \\
& & \nabla^2 V_d (x) = (\mathbf{n} \cdot \nabla)^2\bigg(n_c (x) + \tilde{n} (x) \bigg) \label{dipolar-zng-Vd} \;.
\end{eqnarray}

Having indicated that, we do nonetheless caution the reader that such a procedure has not been shown (here, or elsewhere) to be internally self-consistent, and may thus carry some risks. For example, it is not {\em a priori} possible to ensure that we are not double-counting contributions, or omitting other equally-important further contributions from our equations, a well-known problem in perturbative treatments of quantum gases, see, e.g.~\cite{shi_griffin_98}.

Looking at the postulated form of Eq.~\eqref{dipolar-zng-egpe}, points to the emergence of a dimensionless ratio characterizing the local relative importance of quantum to thermal fluctuations in a dipolar condensate, via
\begin{equation}
{\cal F}({\bf r}) = \frac{\gamma_{QF} \, |\Phi({\bf r})|^3}{2 \, (g-C_{dd}/3) \, \tilde{n}({\bf r})} \;. \label{F-dipolar}
\end{equation}
Such a parameter separates the regimes of sub-dominant quantum fluctuations (${\cal F} \ll 1$) in which our {\em ab initio} formalism emerges naturally, from that when they dominate (${\cal F} \gg 1$). 
We note here that while the net effect will be given by $\int d^3{\bf r}\, {\cal F({\bf r})}$, the inhomogeneous densities (droplets) formed  by the LHY term may in fact imply that its effect cannot be {\em locally} ignored even if $\int d^3{\bf r} {\cal F({\bf r})} \ll 1$ overall in the gas.
As well-known the parameter characterizing their relative importance is $g-C_{dd}/3 = g (1-\epsilon_{dd})$. As $\epsilon_{dd} \rightarrow 1$, as is the case in various dipolar condensate regimes, ${\cal F}({\bf r}) \gg 1$, and so the above postulated contribution $+\gamma_{QF} |\Phi_0|^3\Phi_0$ becomes highly-relevant, even if only locally so.

Alternatively, observation of Eq.~\eqref{F-dipolar} shows that, for $C_{dd}/3 \ll g$, or equivalently $\epsilon_{dd} \ll 1$, and a reasonable intermediate temperature $0 < T < T_c$, the quantum fluctuation term may only pose a minor correction to the overall system properties and dynamics, as expected deeply within the superfluid regime.

The interesting and pertinent question of course remains, whether the above postulated combination of quantum and thermal contributions [Eqs.~\eqref{dipolar-zng-egpe}-\eqref{dipolar-zng-Vd}] (see also related work in Refs.~\cite{politi_pohl_23,he_maucher_2025}), which is known to work well in the respective quantum-dominated and thermal-dominated limits may actually provide valid predictions even in the intermediate regime when ${\cal F} \sim O( 1 )$. In the absence of any other advanced theoretical models addressing this, it may thus be relevant to investigate that numerically, which we defer to a future publication.

\subsection{(Simple Growth) Stochastic Gross-Pitaevskii Equation for Dipolar Condensates}

Alternatively, and similar to the contact-interaction stochastic treatments reviewed earlier, our general set of equations [Eqs.~\eqref{sgpe-full-1}-\eqref{poisson-full}] can also be reduced, upon setting $\Pi^R=\Pi^K=0$, to an effective stochastic equation for the coherent modes of the system in the form
\begin{eqnarray}
i \frac{\partial \Phi_0 (x)}{\partial t} &&= \left[ - \frac{1}{2 m} \nabla^2 + V_{ext} (x) +
C_{dd} V_d(x) +
\left( g - \frac{C_{dd}}{3} \right) \left( n_c (x) + 2 \tilde{n} (x) \right) - \mu \right] \Phi_0 (x) \\
&&-iR \Phi_0 + \xi_1 (x) \;. 
\end{eqnarray}
As discussed earlier [Sec.~\ref{sec:contact}] for the weakly-interacting gas limit, such an equation is not useful to work with on its own, as the term $iR$ is dynamical, encompassing the dynamics of the incoherent modes.

Adapting the earlier arguments, one could make again the assumption of thermal equilibrium, also approximating the Bose-Einstein distribution with a Rayleigh-Jeans one, to reduce the above form to a simpler dynamical  expression for $\Phi_0$, by means of the fluctuation dissipation relation [Eq.~\eqref{iReq}].
Moreover, the Poisson-like equation defining the (in principle dynamical) dipolar potential 
\begin{eqnarray}
&& \nabla^2 V_d (x) = (\mathbf{n} \cdot \nabla)^2\bigg(n_c (x) + \tilde{n} (x) \bigg) \;,
\end{eqnarray}
can be formally integrated to define the potential $V_d$ in the form
\begin{equation}
V_d ({\bf r}) = \frac{1}{4\pi} \int d^3{\bf r'} (\mathbf{n} \cdot \nabla)^2 \, \frac{1}{|{\bf r}-{\bf r'}|} \, \left[ n_c({\bf r'}) +\tilde{n}({\bf r'})  \right] \;.
\end{equation}

Following the above procedure, we may thus obtain a single stochastic equation in the form

\begin{eqnarray}
i \frac{\partial \Phi_0 (x)}{\partial t} = \left( 1 - i \gamma \right) && \left[  - \frac{1}{2 m} \nabla^2 + V_{ext} (x) +
  \frac{C_{dd}}{4\pi} \int d^3{\bf r'} (n \cdot \nabla )^2 \frac{1}{|{\bf r} - {\bf r'}|} \left( n_c({\bf r'}) + \tilde{n}({\bf r'}) \right) \right. \label{dipolar-sgpe-single0a} \\
&&\left. \hspace{0.3cm}+ \left( g - \frac{C_{dd}}{3} \right) \left( n_c (x) + 2 \tilde{n} (x) \right) - \mu \right] \Phi_0 (x) + \xi_1 (x)  \;,
\label{dipolar-sgpe-single0b}
\end{eqnarray}

where the dissipation parameter $\gamma$ is defined here as
\begin{eqnarray}
\gamma &=& i\frac{\beta}{4} \Sigma_{(c)}^K \nonumber \\
&=& \frac{\beta}{2} \left( g -
\frac{C_{dd}}{3} \right)^2
\int \frac{d^3 p_1 d^3 p_2 d^3 p_3}{(2 \pi)^5} \delta
(\varepsilon_{c} + \varepsilon_{\mathbf{p}_1} -
\varepsilon_{\mathbf{p}_2} - \varepsilon_{\mathbf{p}_3}) \delta (\mathbf{p}_c
+ \mathbf{p}_1 - \mathbf{p}_2 - \mathbf{p}_3) \nonumber \\
& & \hspace{2.5cm} \times \bigg[f_1 (1 + f_2) (1 + f_3) + (1 +
f_1) f_2 f_3\bigg] \;. \label{gamma-def}
\end{eqnarray}

One important point that emerges from this equation relates to the explicit appearance of the non-condensate density contribution $\tilde{n}$ in Eqs.~\eqref{dipolar-sgpe-single0a}-\eqref{dipolar-sgpe-single0b}. 
Previously, in the contact-interaction setting, we argued that while any quantitative comparison to experiments strictly requires that contribution to be included, ignoring , to lowest approximation, such a contribution might still give reasonable (but perhaps only qualitative) results, with such treatment possibly also able to model experiments (semi)quantitatively when further supplemented with an appropriately fit $\gamma$ parameter, obtained from experimental measurements. Contrary to that, in the dipolar case we feel it is pertinent to highlight that the situation is more complicated.
The reason for this is that the effective dipolar potential $V_d$, included in its integrated form in Eq.~\eqref{dipolar-sgpe-single0a},
actually depends critically on the thermal density itself, due to the long-range nature of the interactions. As such, it is not {\em a priori} obvious that there exist interesting finite-temperature regimes where ignoring such contributions can be valid. A natural question of course arises how such non-condensate densities could in fact be included in the treatment. Borrowing ideas from earlier treatments, it might be valid in some regimes to consider the thermal cloud density as static, as assumption already implicitly made (through the use of a fluctuation-dissipation relation) in writing the full stochastic equation in the above reduced form. In that case, a typical simulation could, for example, first obtain the self-consistent static thermal equilibrium of a dipolar condensate (e.g. by an extended dipolar Hartree-Fock-like self-consistent method),
and then presumably assume that, under weak perturbations of the thermal cloud, such density remains approximately static\footnote{In the context of the contact-interaction ZNG model, this is referred to as the `static thermal cloud' approximation~\cite{griffin_nikuni_book_09}; note, however, that within such an approximation various dynamical dissipations, e.g. vortex decay, were found to have been underestimated in the case of gases with local interactions}. This could potentially lead to a useful approximate regime under some weakly-perturbed experimental conditions, although one needs to be mindful that the strong coupling induced by the non-local interactions may still lead to non-negligible indirect thermal cloud dynamics via their dipolar-induced interactions with the non-equilibrium condensate.

To avoid such issues altogether, one could make the rather crude approximation of dropping the $\tilde{n}$ term altogether from the above effective near-equilibrium stochastic equation. This will likely produce reasonable qualitative results, but one should not necessarily expect quantitative agreement with experimental findings (in the same way that an effective SGPE description of quasi-1D and quasi-2D gases was shown explicitly to only agree with a broad range of experiments once the beyond classical-field density was taken into consideration, even if within a static approximation).

If one were to nonetheless proceed down that route and thus set $\tilde{n}=0$ in both its occurrences in Eqs.~\eqref{dipolar-sgpe-single0a}-\eqref{dipolar-sgpe-single0b}, and also assume that the parameter $\gamma$ can, for a given experimental configuration, be well-described by an effective constant (i.e.~position- and time-independent) parameter, then one ultimately arrives at

\begin{eqnarray}
i \frac{\partial \Phi_0 (x)}{\partial t} = \left( 1 - i \gamma \right)  && \hspace{-0.3cm} \left[  - \frac{1}{2 m} \nabla^2 + V_{ext} (x) +
 \frac{C_{dd}}{4\pi} \int d^3{\bf r'} (n \cdot \nabla )^2 \frac{1}{|{\bf r} - {\bf r'}|} |\Phi_0({\bf r'})|^2   \right. \label{dipolar-sgpe-single1} \\
&&\left. \hspace{0.3cm}+ \left( g - \frac{C_{dd}}{3} \right) |\Phi_0(x)|^2 - \mu \right] \Phi_0 (x) + \xi_1 (x)  \;,
\label{dipolar-sgpe-single2}
\end{eqnarray}
with correlations 
\begin{equation}
 \langle \xi_1^*(x) \xi_1(x') \rangle = \frac{2}{\beta}\gamma \, \delta(x-x')
\end{equation}

Such an equation corresponds precisely to the  equation used in~\cite{linscott_blakie_14}  to study finite-temperature dipolar condensates, with an {\em ad hoc} (constant) value for $\gamma$ (selected in such work as $\gamma=0.1$, which is considerably larger than typical analytically-estimated values).
Note that, in such work, inclusion of quasiparticle physics was achieved by populating the initial state in the simulations by randomised Bogoliubov particle population.

As  mentioned earlier, the growing interest in dipolar droplets, where the quantum LHY corrections appear to play a dominant part in the system properties, has led to a modified form of the above equation, with the explicit (but somewhat {\em ad hoc} within the premises of such stochastic models) addition of the previously mentioned $\gamma_{QF} |\Phi_0|^3 \Phi_0$ term on the RHS of Eqs.~\eqref{dipolar-sgpe-single1}-\eqref{dipolar-sgpe-single2}, leading to the following extended stochastic GPE:
\begin{align}
i \frac{\partial \Phi_0 (x)}{\partial t} = \left( 1 - i \gamma \right)\hspace{-0.05cm}  &  \left[  - \frac{1}{2 m} \nabla^2 + V_{ext} (x) +
 \frac{C_{dd}}{4\pi} \int d^3{\bf r'} (n \cdot \nabla )^2 \frac{1}{|{\bf r} - {\bf r'}|} |\Phi_0({\bf r'})|^2   \right.& 
 \nonumber \\
&\left. \hspace{0.3cm}+ \left( g - \frac{C_{dd}}{3} \right) |\Phi_0(x)|^2 - \mu \right] \Phi_0 (x) \nonumber \\  &+ \xi_1 (x)  
+\gamma_{QF}(\epsilon_{dd})|\Phi_0|^3 \Phi_0\;.
\label{dipolar-sgpe-general}
\end{align}
 
Such an equation was recently used by Bland {\em et al.}~\cite{bland_poli_22} to study supersolid formation in dipolar condensates, using a constant value for the dissipative $\gamma$ parameter chosen from matching the condensate growth to relevant experimental growth data~\cite{sohmen_politi_21}.

Firstly, we note here that our detailed analysis, when all approximations leading to Eqs.~\eqref{dipolar-sgpe-single1} - \eqref{dipolar-sgpe-single2} are performed, does in fact give an explicit {\em ab initio} expression for the value of $\gamma$ [Eq.~\eqref{gamma-def}], but such an expression needs to be numerically self-consistently evaluated in the context of a particular thermal cloud density approximation, such as the Rayleigh-Jeans distribution in the Hartree-Fock limit. Secondly, it is also pertinent to remind the reader here that the analytical values of $\gamma$ calculated are in fact most likely only reasonable up to an order of magnitude level, rather than an exact, estimate -- despite some interesting evidence~\cite{Rooney13Persistent,mehdi_bradley_21,mehdi_hope_23} in the context of weakly-interacting alkali gases that these analytical expressions for an effective $\gamma$ parameter can even lead to quantitative agreement with experiments. Nonetheless, as in most weakly-interacting gas cases, a more direct route for the choice of the effective parameter $\gamma$ is to choose it such that some dynamical observable is consistent with experimental observations -- and then assume that such value of $\gamma$ remains fixed and produces other system dynamics fairly accurately, so that the equation acquires further predictive powers.
This is precisely the approach used in~\cite{bland_poli_22} and is possibly the most extended treatment one can currently do numerically to model the $T>0$ dipolar gas dynamics, without explicitly solving the S(P)GPE coupled to the quantum Boltzmann equation.

Secondly, and more importantly, all the earlier comments about the self-consistency of first deriving an effective set of equations when thermal fluctuations dominate, and then supplementing it with corrections known to be dominating the quantum-dominated limit, still hold. While we have no reason to argue against trying such an approach, which is indeed the first logical step in that direction in the absence of more concrete numerically-available models, it is important to highlight that there could be competing assumptions/approximations within any formal derivation including both thermal and quantum effects, which could further modify such equation.

This is of course a very challenging question, and remains an open topic of research.  
While acknowledging that
such effective stochastic equations can evidently provide useful insight into dipolar condensate experimental observations~\cite{bland_poli_22}, we nonetheless caution the reader when interpreting novel (or unexpected) results emerging from such simulations.

\subsubsection{Consistency of the LHY Correction Term}

Before closing this stochastic section we add some further remarks that may have some potential relevance here:
A related detailed comparison between different formalisms, most notably between kinetic-type (ZNG) and stochastic-based (SGPE) approaches, has also been elusive in the case of atomic gases with contact interactions\footnote{See, e.g., Chapters  by Griffin and Zaremba on the one hand, and by Wright, Davis and Proukakis on the other discussing such connections in Ref.~\cite{Proukakis13Quantum}.}. 
Some interesting advances were nonetheless made through tailored numerical simulations in the framework of the Projected GPE~\cite{davis_morgan_01,davis_jpb_01} and stochastic GPE settings: specifically, in~\cite{wright_proukakis_11,cockburn_negretti_11} finite-temperature equilibrium data obtained respectively within the PGPE and SGPE were used to extract the condensate contribution through the Penrose-Onsager mode criterion~\cite{Penrose56Bose} (corresponding to the mode with largest eigenvalue). Number-conserving normal and anomalous averages 
of fluctuation operators beyond the numerically-obtained Penrose-Onsager mode
were the constructed, and their relative importance ascertained (e.g. by studying their effect on the `renormalized' equilibrium system chemical potential, which should remain independent of position in a homogeneous system~\cite{wright_proukakis_11}). A key takeaway from such works is that when using simulations with noisy fields (including, e.g., noisy initial conditions in PGPE, dynamical noise in SGPE), anomalous averages of carefully-constructed beyond-condensate number-conserving operators -- which are likely to contain many-body effects, and quantum depletion features related to the LHY correction -- can in fact, at least to some extent, arise naturally within such formalisms. 
In other words, the interpretation of what constitutes the `condensate' within such multi-mode PGPE/SGPE treatments (i.e. contrasting the full $|\Phi_0|^2$ field to the self-consistently determined Penrose-Onsager mode) could imply that the appearance of the LHY correction in the dynamical generalised GPE equations should not necessarily be in the usual form of $\gamma_{QF}|\Phi_0|^3\Phi_0$, as $\Phi_0$ is explicitly  a multi-mode field spanning not only the condensate itself, but also the low-lying modes affected by its presence. Instead, it is plausible that a term with the usual LHY structure should only be written in terms of the corresponding numerically-obtained Penrose-Onsager mode, $\Phi_{PO}$. If such an interpretation were correct, this could somewhat question the premise of using the dipolar SPGPE in the above-stated form of Eq.~\eqref{dipolar-sgpe-general} (without any further amendments). Interestingly, such an issue need not however necessarily arise in the ZNG-like treatment of Sec.~\ref{eQBE}, since there the field $\Phi_0$ refers {\em explicitly} to the condensate mode. While such considerations may in fact have very little effect in practice in the numerical calculations and thus any subsequent comparison to experiments, we nonetheless felt it was pertinent to comment on this here, at least from a fundamental perspective, for completeness.

\section{Conclusions}

In this work we have presented a general non-equilibrium theoretical formalism for modelling Bose gases with a general interatomic potential, which is expected to be valid in a broad range of physical scenarios: while we have explicitly demonstrated their relevance in bridging existing theories and extending beyond them in the context of ultracold Bose gases with contact interactions, and dipolar gas condensates, we anticipate that the more general formulation of our theory (discussed in Appendix A) might in the future also become useful in modelling other systems exhibiting long-range interactions, such as dipolar molecular condensates, Rydberg systems, or indirect excitons.
Our formalism is based on splitting the system into coherent and incoherent degrees of freedom, and including fluctuations of the incoherent modes to all orders, but fluctuations in the coherent modes are only maintained up to second order in the Keldysh q-fields, representing stochastic (statistical) fluctuations. As such, our approach is valid in the domain when thermal fluctuations dominate the system properties; in such a limit, our formalism was shown to generalize a range of established theories, such as the Gross-Pitaevskii-Boltzmann (or ZNG) model, and the stochastic (and projected) Gross-Pitaevskii model -- both of which have found immense success in modelling ultracold atomic experiments with contact interactions.

Motivated by recent discoveries in the context of dipolar gases, where regimes exist in which quantum fluctuations become dominant, helping stabilize the system against collapse, we have given a brief review of how such distinct regime could potentially be handled theoretically. Having an accurate description of the two limiting cases, namely quantum-dominated and thermal-dominated, has enabled us -- following similar lines as a handful of recent works~\cite{bland_poli_22,politi_pohl_23} -- to {\em postulate} a more generalised model which includes {\em both} quantum and thermal fluctuations of the coherent field, and may thus reasonably interpolate between the established limiting cases. In so doing, we have also formulated a potentially-relevant criterion [Eq.~\eqref{F-dipolar}] for the relative importance of quantum and thermal fluctuations in a given system. While the extent of such an extrapolation to the regime where quantum and thermal fluctuations co-exist in comparable measures is unknown, intuition from similar questions in ultracold gases with contact interactions and preliminary evidence from successfully modelling of dipolar gas experiments, provide promising avenues for further exploration with such combined approaches.

Under such an assumption, our final set of equations for non-equilibrium finite-temperature systems exhibiting long-range interactions takes the form of the coupled Eqs.~\eqref{sgpe-full-1}-\eqref{poisson-full}, 
with the explicit inclusion of
an {\em additional} term accounting for the Lee-Huang-Yang quantum correction.
In the case of dipolar gases, such a term would take the usual form of Eq.~\eqref{gamma-qf}, possibly even through its finite-temperature generalized expression also involving thermal densities, which in our treatment can be fully dynamically included through the quantum Boltzmann equation \eqref{zng-full} for the phase-space distribution of the incoherent particles.

While admittedly various open issues remain, we believe this work represents another important step towards the development of systematic studies of non-equilibrium finite-temperature quantum gases with long-range interactions.


\section*{Acknowledgements}
This work was supported by the Leverhulme Trust, Grant no.
RPG-2021-010. We acknowledge insightful discussions with Tom Bland, Ashton Bradley and Mike Garrett, and useful comments by Axel Pelster and Duncan O'Dell.




\appendix

\section{Final Stochastic Equations for a General Interatomic Potential}

Starting from the same Hamiltonian [Eq.~(1)], we discuss here the case of a more general interaction potential of the convenient form
\begin{equation}
\label{interaction}
U (\mathbf{r}, \mathbf{r}') = \alpha_1 \delta (\mathbf{r} - \mathbf{r}') +\alpha_2^2 \mathcal{O}^{- 1} (\mathbf{r}, \mathbf{r}') \;,
\end{equation}
which facilitates expressing the Poisson-like equation in the forms already done in the main paper,
where $\alpha_1$, $\alpha_2$ are yet unspecified constants specific to a particular problem under study.
An expression of this type
enables us to cast the effective action, through 
a Hubbard-Stratonovich transformation in the term involving the interaction $\mathcal{O}^{- 1}$, to the more general form
\begin{equation}
\label{eqini}
S = \int d^4 x \left( i \psi^{\ast} \dot{\psi} + \frac{1}{2
m} \psi^{\ast} \nabla^2 \psi - \frac{\alpha_1}{2} (\psi^{\ast}\psi)^2 - V_{ext} \psi^{\ast}\psi + \frac{1}{2}
V\mathcal{O}V - \alpha_2 V \psi^{\ast}\psi \right) \;,
\end{equation}
where the information on the long-range interactions has now been encoded in the new field $V$ appearing in the effective action after the use of the transformation. Notice that this formulation is equivalent to the initial one. It is possible to see this by taking the equation of motion for $V$:
\begin{equation}
\mathcal{O}V-\alpha_2\psi^\ast \psi=0 \;,
\end{equation}
which gives the solution for $V$ in the form
\begin{equation}
V=\alpha_2\mathcal{O}^{-1}(\psi^\ast \psi) \;,
\end{equation}
where $\mathcal{O}^{-1}$ is understood as the Green's function that makes $\mathcal{O}\mathcal{O}^{-1}=\delta(x-x')$. Plugging back to the action we have that 
\begin{equation}
\label{eqini-app}
S = \int d^4 x \left( i \psi^{\ast} \dot{\psi} + \frac{1}{2
m} \psi^{\ast} \nabla^2 \psi - \frac{\alpha_1}{2} (\psi^{\ast}\psi)^2 - V_{ext} \psi^{\ast}\psi - \frac{1}{2}
\alpha_2^2\psi^\ast \psi\mathcal{O}^{-1}(\psi^\ast \psi) \right) \;,
\end{equation}
which corresponds to the original hamiltonian. The above choice for $U (\mathbf{r}, \mathbf{r}')$ is convenient, as it makes the treatment simpler to manage due to the local expression of the term $V\psi^\ast \psi$. In this sense, both $\alpha_1$ and $\alpha_2$ play the role of interaction strengths, and we treat them as perturbative parameters.

For such an interaction potential, the most general form of the final arising equations becomes
\begin{eqnarray}
& & i \frac{\partial \Phi_0 (x)}{\partial t} = \left( - \frac{1}{2 m} \nabla^2 +
V_{ext}(x) + V_c(x) \right) \Phi_0 (x)\nonumber\\
& & \hspace{1.8cm} - i R(x) \Phi_0 (x) + \xi_1(x) \nonumber \\
& & \hspace{1.8cm}  - 2 \alpha_1 \int d^4 x' \Pi^R (x', x) V_{nc} (x') \Phi_0 (x) +
\alpha_1 \xi_2(x) \Phi_0(x) \;, \label{condensateeq-app}\\
& & \mathcal{O}V_d (x) = \alpha_2 (n_c (x) + \tilde{n} (x)) - \alpha_2
\int d^4 x' \Pi^R (x', x) V_{nc} (x') + \frac{1}{2} \alpha_2 \xi_2(x) \;, \label{vcleq-app}\\\
& & \frac{\partial f}{\partial t} + \frac{\mathbf{p}}{m} \cdot\nabla f - \nabla \bigg(V_{ext}(x)
+ V_{nc}(x)\bigg) \cdot \nabla_{\mathbf{p}} f =
\frac{1}{2} (I_a + I_b) \;, \label{particleeq-app}
\end{eqnarray}

where we recall that the mean field potentials for the coherent and non-coherent parts are respectively
\begin{eqnarray}
V_c (x) &=& \alpha_2 V_d (x) + \alpha_1 (n_c (x) + 2 \tilde{n} (x)) \\
V_{nc} (x) &=& \alpha_2 V_d (x) + 2 \alpha_1 (n_c (x) +
\tilde{n} (x))
\end{eqnarray}
and the number densities for the coherent and non-coherent parts are
\begin{equation}
n_c=|\Phi_0|^2, \quad \tilde{n}=\int \frac{d^3 p}{(2\pi)^3}f(x,\mathbf{p}) \;.
\end{equation}

The terms in the right hand side of \eqref{particleeq-app} corresponding to collisional terms of this Boltzmann equation are given by
\begin{eqnarray}
I_a &=& 4 \alpha_1^2 n_c \int \frac{d^3 p_1 d^3 p_2 d^3 p_3}{(2 \pi)^2}
\delta (\varepsilon_{\mathbf{q}} + \varepsilon_{\mathbf{p}_1} -
\varepsilon_{\mathbf{p}_2} - \varepsilon_{\mathbf{p}_3}) \delta
(\mathbf{p}_2 - \mathbf{p}_1 - \mathbf{q} + \mathbf{p}_3)\nonumber\\
& & \hspace{1.2cm} \times (\delta (\mathbf{p}_1 - \mathbf{p}) - \delta (\mathbf{p}_2 -
\mathbf{p}) - \delta (\mathbf{p}_3 - \mathbf{p})) ((1 + f_1) f_2 f_3 -
f_1 (1 + f_2) (1 + f_3)) \nonumber\\
I_b &=& 4 \alpha_1^2 \int \frac{d^3 p_2 d^3 p_3 d^3 p_4}{(2 \pi)^5} \delta
(\varepsilon_{\mathbf{p}_3} + \varepsilon_{\mathbf{p}_4} -
\varepsilon_{\mathbf{p}_2} - \varepsilon_{\mathbf{p}}) \delta
(\mathbf{p} + \mathbf{p}_2 - \mathbf{p}_3 - \mathbf{p}_4) \nonumber\\
& & \times [f_3 f_4 (f + 1) (f_2 + 1) - f f_2 (f_3 + 1) (f_4 + 1)] \label{eqcolls-app}
\end{eqnarray}
From the explicit expression for $R$ 
we have that
\begin{eqnarray}
R &=& \alpha_1^2
\int \frac{d^3 p_1 d^3 p_2 d^3 p_3}{(2 \pi)^5} \delta
(\varepsilon_{\mathbf{q}} + \varepsilon_{\mathbf{p}_1} -
\varepsilon_{\mathbf{p}_2} - \varepsilon_{\mathbf{p}_3}) \delta (\mathbf{q} + \mathbf{p}_1
- \mathbf{p}_3 - \mathbf{p}_2) \nonumber\\
& & \times \bigg[ f_1 (1 + f_2) (1 + f_3) - (1 + f_1) f_2 f_3 \bigg]\nonumber\\
&=& \frac{1}{4 n_c} \int \frac{d^3 p}{(2 \pi)^3} I_a \;. \label{rdef2-app}
\end{eqnarray}

Also,
\begin{equation}
\Pi^R (x, \mathbf{k}) = \int \frac{d^3 p_1 d^3 p_2}{(2 \pi)^3}
\frac{1}{\varepsilon_{\mathbf{k}} + \varepsilon_{\mathbf{p}_2} -
\varepsilon_{\mathbf{p}_1} + i \sigma} \delta (\mathbf{k} + \mathbf{p}_2
- \mathbf{p}_1) \bigg[f_1 (1 + f_2) - f_2 (1 + f_1)\bigg] \;,
\end{equation}

and the noise terms $\xi_1$ and $\xi_2$ satisfy the correlations
\begin{eqnarray}
\langle \xi_1^{\ast} (x) \xi_1 (x') \rangle &=& \frac{i}{2} \Sigma_{(c)}^K (x)\delta(x-x')\\
\langle \xi_2 (x) \xi_2 (x') \rangle&=&- 2 i \Pi^K (x, x') \label{realcorrelators1-app}
\end{eqnarray}
where we remember that
\begin{eqnarray}
\Sigma_{(c)}^K(x) &=& - 2 i \alpha_1^2
\int \frac{d^3 p_1 d^3 p_2 d^3 p_3}{(2 \pi)^5} \delta
(\varepsilon_{\mathbf{q}} + \varepsilon_{\mathbf{p}_1} -
\varepsilon_{\mathbf{p}_2} - \varepsilon_{\mathbf{p}_3}) \delta (\mathbf{q}
+ \mathbf{p}_1 - \mathbf{p}_2 - \mathbf{p}_3) \nonumber\\
& & \hspace{1.0cm} \times \bigg[f_1 (1 + f_2) (1 + f_3) + (1 +
f_1) f_2 f_3\bigg] \;, \label{sigkdef-app}\\
\Pi^K (x, \mathbf{k}) &=&  i \int \frac{d^3 p_1 d^3 p_2}{(2 \pi)^2} \delta
(\varepsilon_{\mathbf{k}} + \varepsilon_{\mathbf{p}_2} -
\varepsilon_{\mathbf{p}_1}) \delta (\mathbf{k} + \mathbf{p}_2 -
\mathbf{p}_1) \bigg[f_1 (1 + f_2) + f_2 (1 + f_1)\bigg] \;. \label{pikdef-app}
\end{eqnarray}

These equations constitute the main equations for a bosonic system composed by a set of coherent and non-coherent particles interacting with an additional potential that we have encoded in $V_d$.

The dipolar case explicitly considered in this manuscript arises as a special case of this, as do other common cases mentioned below based on their implicit choices:
\begin{enumerate}
\item[(a)] 
 Dipolar Gases: 
 \begin{equation}
\hspace{-2.0cm} \alpha_1=g-\frac{C_{dd}}{3}, \quad \quad \quad  \alpha_2=C_{dd}, \quad \quad \quad {\rm and} \quad \quad \quad
\mathcal{O}=C_{dd}\,\frac{\nabla^2}{(\mathbf{n} \cdot \nabla)^2} \;.
\end{equation}
\item[(b)] s-wave Interacting Gases: 
\begin{equation}
\hspace{-7.8cm} \alpha_1=g, \quad \quad \quad \quad \quad 
\alpha_2 = 0 \;.
\end{equation}
\item[(c)] Cosmological model  (see companion paper~\cite{proukakis_rigopoulos_arxiv_24}):
\begin{equation}
\hspace{-1.5cm}
\alpha_1=g, \quad \quad \quad \quad \quad
\alpha_2=m, \quad \quad \quad \quad 
{\rm and} \quad \quad \quad \quad 
\mathcal{O}=\frac{1}{4 \pi G}\, \nabla^2 \;.
\end{equation}

\end{enumerate}

\section{Non-local operator properties }\label{appNonLocal}

Our equations include a non-local operator $\frac{1}{(\mathbf{n}\cdot\nabla)^2}$, which we stated corresponds to the inverse of the operator $(\mathbf{n}\cdot \nabla)^2$ via the relation
\begin{equation}
\label{nonlocalop}
\frac{1}{(\mathbf{n} \cdot \nabla)^2} (\mathbf{n} \cdot \nabla)^2 = 1.
\end{equation}

These non-local operators has been studied commonly in a four dimensional space-time setup in the context of Very Special Relativity \cite{Cohen2006ky} and also in the context of non-covariant gauges \cite{Leibbrandt1987qv}, as the Light Cone Gauge in QCD (for example in \cite{Das2004wm}). 
In practical terms, in a four dimensional space-time these operators has been treated typically in Fourier space, which is the easiest way to perform computations with them. 

As an operator, it acts on functions, so, in an operational way, let us take an arbitrary function of space $f(\mathbf{r})$.
To confirm such a statement, here we prove that
\begin{equation}
\frac{1}{(\mathbf{n} \cdot \nabla)^2} (\mathbf{n} \cdot \nabla)^2 f(\mathbf{r}) = f(\mathbf{r}) \;.
\end{equation}
The best way to work with this class of operators is in Fourier space. There, it is easy to show that
\begin{equation}
(\mathbf{n} \cdot \nabla)^2 f(\mathbf{r})=\int d^3 k (\mathbf{n} \cdot \nabla)^2 f(k)e^{-i \mathbf{k}\cdot\mathbf{r}}= -\int d^3 k (\mathbf{n}\cdot\mathbf{k})^2 f(k)e^{-i \mathbf{k}\cdot\mathbf{r}}
\end{equation}
Now, we use the operator $\frac{1}{(\mathbf{n} \cdot \nabla)^2}$. Therefore,
\begin{equation}
\frac{1}{(\mathbf{n} \cdot \nabla)^2} (\mathbf{n} \cdot \nabla)^2 f(\mathbf{r}) = -\frac{1}{(\mathbf{n} \cdot \nabla)^2} \int d^3 k (\mathbf{n}\cdot\mathbf{k})^2 f(k)e^{-i \mathbf{k}\cdot\mathbf{r}} 
\end{equation}

We will employ a formal definition for this non-local operator \cite{Alfaro:2013uva,Bufalo:2020cst}:
\begin{equation}
\frac{1}{(\mathbf{n} \cdot \nabla)^2} = \int^{\infty}_0 d \alpha e^{-
\alpha (\mathbf{n} \cdot \nabla)^2} = \int^{\infty}_0 d \alpha (1 - \alpha
(\mathbf{n} \cdot \nabla)^2 + \cdots +)
\end{equation}

Hence,
\begin{equation}
\frac{1}{(\mathbf{n} \cdot \nabla)^2} (\mathbf{n} \cdot \nabla)^2 f(\mathbf{r}) = -\int d^3 k \int^{\infty}_0 d \alpha  (1 + \alpha
(\mathbf{n} \cdot \mathbf{k})^2 + \cdots +)(\mathbf{n}\cdot\mathbf{k})^2 f(k)e^{-i \mathbf{k}\cdot\mathbf{r}}
\end{equation}
and recognising that
\begin{equation}
-\frac{1}{(\mathbf{n}\cdot\mathbf{k})^2}=\int^{\infty}_0 d \alpha  (1 + \alpha
(\mathbf{n} \cdot \mathbf{k})^2 + \cdots +)
\end{equation}
we obtain
\begin{equation}
\frac{1}{(\mathbf{n} \cdot \nabla)^2} (\mathbf{n} \cdot \nabla)^2 f(\mathbf{r}) = \int d^3 k f(k)e^{-i \mathbf{k}\cdot\mathbf{r}}=f(\mathbf{r})
\end{equation}
which thus facilitates our writing the Poisson-like equation in the forms already done in the main paper.

\section{Derivation of LHY term from equations of motion }\label{appLHY}


Let us summarize the standard way to obtain the LHY correction. We remark again, this derivation doesn't follow from the Schwinger-Keldysh formalism employed to obtain our set of equations. For this case we start from the equation of motion for the condensate (different from the Schwinger-Keldysh case, where we worked from the action):
\begin{equation}
\label{initialeqforlhy}
i \partial_t \Phi = - \frac{\hbar^2}{2 m} \nabla^2 \Phi + V_{ext} \Phi
+ \int d^3 r' V_{int} (r - r') | \Phi (r') |^2 \Phi
\end{equation}
where we have set for simplicity
\begin{equation}
V_{int}(r-r') = g \delta (r - r') + U_{dd} (r - r')
\end{equation}

We split  $\Phi (r, t) = e^{- i \mu t} (\Phi_0 (r) + \eta (r, t))$ where the term $\eta$ corresponds to the quantum fluctuations in the condensate, and we are considering here $\Phi_0$ slower than the quantum fluctuations and we can consider that the kinetic term for the condensate is zero.
With this splitting, we have two equations, the order zero, which gives an expression for the chemical potential:
\begin{equation}
\label{zerothchem}
0 = - \mu \Phi_0 + V_{ext}
\Phi_0 + \int d^3 r' V_{int} (r - r') | \Phi_0 (r') |^2 \Phi_0
\end{equation}
and the equation to order one in $\eta$:
\begin{equation}
\label{order1eta}
i \partial_t \eta = - \frac{\hbar^2}{2 m} \nabla^2 \eta - \mu \eta +
V_{ext} \eta + \int d^3 r' V_{int} (r - r') (\eta^{\ast} (r')
\Phi_0 (r') + \Phi^{\ast}_0 (r') \eta (r')) \Phi_0 + \int d^3 r'
V_{int} (r - r') | \Phi_0 (r') |^2 \eta
\end{equation}

We will work in this last equation, considering the Bogoliubov transformation
\begin{equation}
\label{etabogoliubov}
\eta = \underset{k}{\sum} (u_k (r) \chi_k (t) + v^{\ast}_k (r) \chi^{\ast}_k
(t))
\end{equation}
where the new objects $u_k$ and $v_k$ satisfy $\int d^3 r (u_k^{\ast} (r) u_l (r) - v_k^{\ast} (r) v_l (r)) =
\delta_{k l}$.
Using this new variables our equation \eqref{order1eta} can be written in a matricial way as
\begin{equation}
\varepsilon_k \left(\begin{array}{c}
  u_k (r)\\
  - v_k (r)
\end{array}\right) = \int d^3 r' \left(\begin{array}{cc}
  H_c + \Phi^{\ast}_0 (r') V_{int} (r - r') \Phi_0 (r) & \Phi_0 (r')
  V_{int} (r - r') \Phi_0 (r)\\
  \Phi^{\ast}_0 (r') V_{int} (r - r') \Phi^{\ast}_0 (r) & H_c +
  \Phi^{\ast}_0 (r) V_{int} (r - r') \Phi_0 (r')
\end{array}\right) \left(\begin{array}{c}
  u_k (r')\\
  v_k (r')
\end{array}\right)
\end{equation}
where we defined
\begin{equation}
H_c = \left( \frac{\hbar^2 k^2}{2 m} - \mu + V_{ext} \right) \delta (r
- r') + V_{int} (r - r') | \Phi_0 (r') |^2
\end{equation}

Due to the non-locality of this equation, the solution is complicated. For this, we can approximate the non-local terms as
\begin{eqnarray}
\int d^3 r' \Phi_0 (r') V_{int} (r - r') \Phi_0 (r) u_k (r') \approx
u_k (r) \Phi_0 (r) \Phi_0 (r) V_{int} (k)\\
\int d^3 r' \Phi^{\ast}_0 (r') V_{int} (r - r') \Phi_0 (r) u_k (r')
\approx u_k (r) | \Phi_0 (r) |^2 V_{int} (k)
\end{eqnarray}
where we use that the Fourier transform of the interacting part is given by
\begin{equation}
\label{vintk}
V_{int} (k) = g [1 + \epsilon_{dd} (3 \cos^2 \theta
- 1)]
\end{equation}
Doing that, it is possible to show that the energy is given by
\begin{equation}
\varepsilon_k = \sqrt{\frac{\hbar^2 k^2}{2 m} \left( \frac{\hbar^2 k^2}{2 m}
+ 2 g | \Phi_0 |^2 [1 + \epsilon_{dd} (3 \cos^2 \theta - 1)] \right)}
\end{equation}
and using the matricial equation with the zeroth order equation \eqref{zerothchem} to eliminate the chemical potential we obtain that
\begin{eqnarray}
| u_k |^2 = \frac{1}{2 \varepsilon_k} \left( \frac{\hbar^2 k^2}{2 m} +
V_{int} (k) | \Phi_0 |^2 \right) + \frac{1}{2} \label{uk2}\\
| v_k |^2 = \frac{1}{2 \varepsilon_k} \left( \frac{\hbar^2 k^2}{2 m} +
V_{int} (k) | \Phi_0 |^2 \right) - \frac{1}{2} \label{vk2}
\end{eqnarray}

On the other hand it is possible to show that from the definition of the transformation \eqref{etabogoliubov} we can have that
\begin{equation}
\langle \eta^{\ast} \eta \rangle = \underset{k}{\sum} ((| u_k |^2 + | v_k
|^2) \langle \chi^{\ast}_k \chi_k \rangle + | v_k |^2)
\end{equation}
Due to the operator nature of the objects $\chi$, we have got this last term that remains at zero temperature. Thus, neglecting the first one, using \eqref{vk2} and passing the sum to an integral, at zero temperature we have a correction to the condensate given by
\begin{equation}
\langle \eta^{\ast} \eta \rangle = \frac{1}{2} \int \frac{d^3 k}{(2 \pi)^3}
\left[ \frac{\frac{\hbar^2 k^2}{2 m} + V_{int} (k) | \Phi_0
|^2}{\sqrt{\frac{\hbar^2 k^2}{2 m} \left( \frac{\hbar^2 k^2}{2 m} + 2
V_{int} (k) | \Phi_0 |^2 \right)}} - 1 \right]
\end{equation}
We will work this integration in spherical coordinates. Therefore, $d^3k=k^2\sin\theta d\varphi d\theta dk$ and since $V_{int}(k)$ depends on the angle (see eq. \eqref{vintk}), which is part of the integration, we need to be careful. We do the integrals in $\varphi$ and $k$ and doing the change of variables $\cos\theta=u$ we finally obtain that
\begin{equation}
\label{etastareta}
\langle \eta^{\ast} \eta \rangle = \frac{1}{3 \pi^2} \left( \frac{m
g}{\hbar^2} \right)^{3 / 2} \mathcal{Q}_3 (\epsilon_{dd}) | \Phi_0 |^3
+ \text{Infinite part}
\end{equation}
where we have defined the auxiliary function
\begin{equation}
\mathcal{Q}_l (\epsilon_{dd}) = \int^1_0 d u (1 - \epsilon_{dd}
+ 3 \epsilon_{dd} u^2)^{l / 2}
\end{equation}
for any integer $l$.

We discard the infinite contribution since it can be eliminated via renormalization. Thus we have obtained the correction at zero temperature of the density. The same can be done in the energy, considering the splitting in the eq. \eqref{initialeqforlhy} up to order two in $\eta$ and after averaging, take the same approximation for the non-local terms used before and using \eqref{uk2} and \eqref{vk2} together with with $g = \frac{4 \pi \hbar^2 a}{m}$ we get an explicit expression for the LHY correction
\begin{equation}
\Delta E= \frac{32}{3} g \sqrt{\frac{a^3}{\pi}} \mathcal{Q}_5 (\epsilon_{dd})|
\Phi_0 |^3\Phi_0
\end{equation}

\bibliographystyle{unsrt}
\bibliography{Dipolar}
 
\end{document}